\documentclass[11pt,english]{elsarticle}
\usepackage[T1]{fontenc}
\usepackage[latin9]{inputenc}
\usepackage{geometry}
\usepackage{array}
\usepackage{booktabs}
\usepackage{units}
\usepackage{multirow}
\usepackage{amsmath}
\usepackage{amsthm}
\usepackage{amssymb}
\usepackage{graphicx}
\usepackage{setspace}
\makeatletter

\providecommand{\tabularnewline}{\\}

\numberwithin{equation}{section}
\numberwithin{figure}{section}
\theoremstyle{plain}
\newtheorem{thm}{\protect\theoremname}
\theoremstyle{plain}
\newtheorem{prop}[thm]{\protect\propositionname}

\usepackage{placeins}  
\usepackage{hyperref}
\usepackage[T1]{fontenc}
\usepackage{graphicx}
\newcommand\myt{\vspace{3mm} \setlength{\tabcolsep}{3pt} \renewcommand{\arraystretch}{0.8}}
\newcommand\mytt{\vspace{4mm} \setlength{\tabcolsep}{4pt} \renewcommand{\arraystretch}{0.9}}
\newcommand\myttt{\vspace{3mm} \setlength{\tabcolsep}{1pt} \renewcommand{\arraystretch}{0.7}}

\makeatother

\usepackage{babel}
\providecommand{\propositionname}{Proposition}
\providecommand{\theoremname}{Theorem}

\begin{document}

\begin{frontmatter}{}

\title{Moving average options: Machine Learning and Gauss-Hermite quadrature
for a double non-Markovian problem}

\author{Ludovic Goudenège}

\ead{ludovic.goudenege@math.cnrs.fr}

\address{Fédération de Mathématiques de CentraleSupélec - CNRS FR3487, France}

\author{Andrea Molent}

\ead{andrea.molent@uniud.it }

\address{Dipartimento di Scienze Economiche e Statistiche, Università degli
Studi di Udine, Italy}

\author{Antonino Zanette}

\ead{antonino.zanette@uniud.it }

\address{Dipartimento di Scienze Economiche e Statistiche, Università degli
Studi di Udine, Italy}

\cortext[cor3]{Corresponding author}
\begin{abstract}
Evaluating moving average options is a tough computational challenge
for the energy and commodity market as the payoff of the option depends
on the prices of a certain underlying observed on a moving window
so, when a long window is considered, the pricing problem becomes
high dimensional. We present an efficient method for pricing Bermudan
style moving average options, based on Gaussian Process Regression
and Gauss-Hermite quadrature, thus named GPR-GHQ. Specifically, the
proposed algorithm proceeds backward in time and, at each time-step,
the continuation value is computed only in a few points by using Gauss-Hermite
quadrature, and then it is learned through Gaussian Process Regression.
We test the proposed approach in the Black-Scholes model, where the
GPR-GHQ method is made even more efficient by exploiting the positive
homogeneity of the continuation value, which allows one to reduce
the problem size. Positive homogeneity is also exploited to develop
a binomial Markov chain, which is able to deal efficiently with medium-long
windows. Secondly, we test GPR-GHQ in the Clewlow-Strickland model,
the reference framework for modeling prices of energy commodities.
Finally, we consider a challenging problem which involves double non-Markovian
feature, that is the rough-Bergomi model. In this case, the pricing
problem is even harder since the whole history of the volatility process
impacts the future distribution of the process. The manuscript includes
a numerical investigation, which displays that GPR-GHQ is very accurate
and it is able to handle options with a very long window, thus overcoming
the problem of high dimensionality.
\end{abstract}

\end{frontmatter}{}

Declarations of interest: none.

\section{Introduction}

In this manuscript, we are interested in path-dependent options whose
payoff depends on the average of the prices of a certain underlying,
observed in a sliding window. Options on the moving average price
are widespread in the commodities sector and in particular in the
gas and oil market. In this context, these options are known as ``swing
options'' (see, e.g. \citet{bernhart2011}) and allow the holder
to buy a certain amount of gas or oil at the average price observed
in a certain window, in order to compensate for possible anomalous
peaks in prices: for example, the average of the daily closing prices
of the last month or of the last week. Furthermore, moving average
options are also popular in corporate finance, where they are used
to protect a company from hostile acquisitions (see, e.g. \citet{dai2010}).

The payoff of moving average options depends on the average of a fixed
number of the last observed daily closing prices. Usually, the moving
window can include from a few days to a few tens (\citet{kao2003},
\citet{bernhart2011}). These options are similar to Asian options,
whose payoff depends on the average price of the underlying, calculated
from the inception time to the exercise time. However, the valuation
of Asian options is relatively simple since, under standard assumptions,
the average over the entire time interval and the underlying price
define a Markov process. On the contrary, the sliding window feature
of the moving average options makes their evaluation much more complicated
as the moving average and the underlying do not define a Markov process:
the price of the option is a function of every single value that is
used to compute the average, and therefore a function of a potentially
large number of variables. 

Pricing American-style moving average options has garnered the attention
of several authors. The approaches proposed in this regard are of
three types essentially. The first group of methods includes Longstaff-Schwartz
type approaches, which is a state-of-the-art approach in the energy
sector (see, e.g. \citet{nadarajah2017}). One of the first attempts
in this field is due to \citet{bilger2003}, which uses a standard
Longstaff-Schwartz method. \citet{grau2008} improves the numerical
efficiency of such an approach by using a sparse polynomial basis
for the regression. \citet{broadie2008} propose to use polynomials
of the underlying index and the average, as well as to exploit two
variance reduction techniques to improve convergence. \citet{bernhart2011}
propose a method based on Laguerre polynomial approximation to price
the continuously monitored moving average options. \citet{dirnstorfer2013}
exploit sparse grid basis functions based on polynomials or piecewise
linear functions. More recently, \citet{lelong2019} introduces a
Longstaff-Schwartz type approach in which the standard least-square
regression is replaced by a Wiener chaos expansion.

The second group includes techniques based on partial differential
equations (PDE). \citet{dai2010} introduce an algorithm for pricing
discretely monitored moving average barrier options based on the resolution
of a PDE and obtain the price of continuously monitored option by
Richardson's extrapolation. \citet{federico2015} study from a theoretical
perspective stochastic delay differential equations which may be used
for computing the price of moving average options.

The third group involves techniques that rely on lattices, which is
common approach to price options on averages (see, e.g. \citet{costabile2011},
\citet{gambaro2020}). \citet{kao2003} present pricing algorithms
based on the CRR binomial model for geometric and arithmetic moving-average
type options. \citet{xu2013} propose a sampling strategy that improves
willow trees. \citet{lu2017} propose two willow tree methods for
pricing European-style and American-style moving average barrier options. 

Finally, it is worth mentioning the works of \citet{dong2019,dong2021}
who study a particular type of swing contracts that include a moving
average feature, by means of a two-dimensional trinomial tree and
least squares Monte Carlo methods. To conclude this review, we also
mention the research on the dynamic hedging of moving average options
of \citet{warin2012}.

The previously mentioned works consider moving average options with
a moving window that includes at most 10 observations (see \citet{dirnstorfer2013},
\citet{bernhart2011}, \citet{lelong2019}). The difficulty in going
beyond this number of observations is due to the computational complexity
of the problem. When the average is computed on many observations,
the evaluation of American-style moving average options becomes a
high dimensional problem which suffers from the well-known curse of
dimensionality.

In the last few years, Gaussian Process Regression (GPR), a Machine
Learning technique that allows estimations from a certain number of
observed values, has recently been used to face high-dimensional problems
in finance. In this regard, we mention the work of \citet{ludkovski2018},
who evaluates Bermudan options by fitting the continuation values
through GPR and \citet{de2018}, that exploit GPR to speed up derivative
pricing by using an online-offline approach. More recently, \citet{goudenege2020}
employ GPR for pricing American options on a basket of assets following
multi-dimensional Black-Scholes dynamics.

In this paper, we show how to approach the evaluation of high-dimensional
Bermudan moving average options by using GPR and Gauss-Hermite quadrature.
Specifically, we propose a method, called GPR-GHQ, that works by moving
backward in time. At each time step, the continuation value is computed
only for some quasi-random data points, which represent some possible
observed values of the underlying, by using Gauss-Hermite quadrature
(see, e.g. \citet{judd1998}). Then, following the same approach of
\citet{ludkovski2018} and \citet{goudenege2020}, we exploit GPR
to learn the continuation value of the moving average option from
the few observed values. First of all, we focus on the Black-Scholes
model. In this particular case, the continuation value is a positive
homogeneous function of observed underlying values and such a property
can be exploited to reduce by one the dimension of the problem. Moreover,
such a feature also allows us to define an efficient pricing model
based on a binomial Markov chain, which turns out to be particularly
efficient for medium-high dimensional options. Since moving average
options are also used in the energy markets, we also test the GPR-GHQ
method in the Clewlow-Strickland model for prices of energy commodities,
which is able to match the forward curve and also provides mean reversion.
Recently \citet{alfeus2020} show that the volatility in the commodity
market is rough, therefore we also consider the rough-Bergomi model,
introduced by \citet{bayer2016}, which is a promising rough volatility
model in quantitative finance. This latter model is particularly interesting
as the underlying is as a process with fractional stochastic volatility.
Therefore, in this case, we face a double non-Markovian option pricing
problem: the absence of Markov property is due both to the underlying
process, which has volatility with memory, and to the moving average,
which is not a Markov process even in the simplest stochastic models,
as it depends on all underlying values included in the moving window.
We present numerical results for all the three stochastic models mentioned
above and we perform an empirical convergence analysis to show that
the proposed method outperforms the standard Longstaff-Schwartz algorithm
and it is very accurate and efficient in handling high dimensional
moving average options. 

The remainder of the paper is organized as follows. In Section 2 we
present the considered stochastic models. In Section 3 we present
moving average options. In Section 4 we outline the main features
of the GPR-GHQ method. In Section 5 we discuss the binomial chain
method. In Section 6 we present and discuss the results of the numerical
simulations. In Section 7 we conclude.

\section{The stochastic models}

In this paper we consider three stochastic models, namely the Black-Scholes
model, the Clewlow-Strickland model and the rough-Bergomi model. In
order to fix the notation, we report models dynamics under a risk
neutral probability $\mathbb{P}$. Simulation procedure for the rough-Bergomi
model is outlined in the  \ref{sec:Simulation-of-Stochastic}, while
for the other two models, we refer the interested reader, e.g., to
\citet{fusai2007}. 

\subsection{The Black-Scholes model\label{subsec:The-Black-Scholes-model}}

The Black-Scholes model is widely recognized as one of the most important
models in finance. It models the dynamics of a stock price by the
following stochastic differential equation (SDE):

\begin{equation}
\frac{dS_{t}}{S_{t}}=r\,dt+\sigma\,dB_{t},\label{eq:BS}
\end{equation}
with $r$ the risk free interest rate, $\sigma$ the volatility and
$\left(B_{t}\right)_{t\in\left[0,T\right]}$ a Brownian motion. 

\subsection{The Clewlow-Strickland model}

The Clewlow-Strickland model \citep{clewlow1999} is the standard
model for commodity price dynamics. Let $S_{t}$ be the spot price
of a commodity and let $F\left(t,T\right)$ denote its forward price
at time $t\in\left[0,T\right]$ with maturity $T$. The model assumes
that $F\left(t,T\right)$ is the solution of the following SDE
\begin{equation}
\frac{dF(t,T)}{F(t,T)}=\sigma e^{-\alpha(T-t)}dB_{t},\label{eq:CS1}
\end{equation}
with $\alpha$ and $\sigma$ positive constants and $\left(B_{t}\right)_{t\in\left[0,T\right]}$
a Brownian motion. Starting from (\ref{eq:CS1}), one can prove that
the spot price is given by the following relation
\begin{align}
S_{t} & =F(0,t)\exp\left[\frac{\sigma^{2}}{4\alpha}\left(e^{-2\alpha t}-1\right)+\int_{0}^{t}\sigma e^{-\alpha(t-u)}dB_{u}\right].
\end{align}

\subsection{The rough-Bergomi model}

The rough-Bergomi model is a non-Markovian model, recently introduced
by \citet{bayer2016}, that provides stochastic volatility with memory
and it is appreciated as it generates a realistic term structure of
at-the-money volatility skew. The model is described by the following
SDE:
\[
\begin{cases}
dS_{t} & =rS_{t}dt+\sqrt{V_{t}}S_{t}dB_{t}^{1}\\
V_{t} & =\xi_{0}\left(t\right)\exp\left(\eta\widetilde{B}_{t}^{H}-\frac{1}{2}\eta^{2}t^{2H}\right),
\end{cases}
\]
with $r$ the risk free interest rate, $\eta$ a positive parameter,
$H\in\left]0,1\right[$ the Hurst parameter and $\xi_{0}\left(t\right)$
a deterministic function that models the forward variance curve. The
process $B_{t}^{1}$ is a Brownian motion, whereas $\widetilde{B}_{t}^{H}$
is a Riemann-Liouville fractional Brownian motion, a non-Markovian
process that can be expressed as 
\begin{equation}
\widetilde{B}_{t}^{H}=\sqrt{2H}\int_{0}^{t}\left(t-s\right)^{H-\frac{1}{2}}dB_{t}^{2},
\end{equation}
 with $B_{t}^{2}$ a Brownian motion and $\rho$ the instantaneous
correlation coefficient between $B_{t}^{1}$ and $B_{t}^{2}$. 

\section{\label{sec:Moving-Average-options}Moving average options}

We consider a time interval $\left[0,T\right]$ and a stochastic process
$\left(S_{t}\right)_{t\in\left[0,T\right]}$ that models a market
index, for example the price of a stock or of a commodity. Let us
suppose that there are $N$ trading dates $t_{1},\dots t_{N}$ in
$\left]0,T\right]$ with $t_{n}=\nicefrac{nT}{N}$, so that $t_{N}=T$.
Let $n_{1},n_{2}$ be two integers with $0\leq n_{1}\leq n_{2}\leq N$,
and let $A_{n_{1}}^{n_{2}}$ represent the average closing value of
the process $S$ form $t_{n_{1}}$ to $t_{n_{2}}$, that is
\[
A_{n_{1}}^{n_{2}}=\frac{1}{n_{2}-n_{1}+1}\sum_{j=n_{1}}^{n_{2}}S_{t_{j}}.
\]
We stress out that $A_{n}^{n}=S_{t_{n}}$. The payoff of a moving
average option at time $t_{n}$ is given by
\begin{equation}
\Psi\left(S_{t_{n-M+1}},S_{t_{n-M+2}},\dots,S_{t_{n}}\right)=\max\left(0,S_{t_{n}}-A_{n-M+1}^{n}\right),
\end{equation}
 with $M$ the number of observed underlying values included in the
average. Please observe that, since the first available underlying
value is $S_{0}$, the payoff function can not be evaluated before
time $t_{M-1}$.

We are interested in Bermudan options, which can be exercised at any
time step $t_{n}$, for $n=M,\dots,N$. It is worth underlying that,
following \citet{bernhart2011} and \citet{lelong2019}, the first
time the option can be exercises is $t_{M}$, but other choices are
possible.

Although this type of option may have the appearance of an Asian call
with a floating strike, there is a very important difference. Unlike
what happens for Asian options, the pair $\left(S_{t_{n}},A_{n-M+1}^{n}\right)_{n=M,\dots,N}$
does not define a Markov process, even if $\left(S_{t_{n}}\right)_{n=M,\dots,N}$
is a Markov process. In fact, the updating rule for $A$ is given
by
\begin{equation}
A_{\left(n+1\right)-M+1}^{\left(n+1\right)}=A_{n-M+1}^{n}+\frac{S_{t_{n+1}}-S_{t_{n-M+1}}}{M},
\end{equation}
so that the distribution law of the couple $\left(S_{t_{n+1}},A_{n+1-M+1}^{n+1}\right)$
does not depend only on $\left(S_{t_{n}},A_{n-M+1}^{n}\right)$ but
also on $S_{t_{n-M+1}}$. That is because the averaging window moves
with time and so the oldest underlying value included in the average
has to be removed from the average to leave room for the newest underlying
value. By exploiting a similar reasoning, one can show that the distribution
law of $\left(S_{t_{n+2}},A_{\left(n+2\right)-M+1}^{\left(n+2\right)}\right)$
depends on $S_{t_{n-M+2}}$ and so on, so that all the $M$ values
$S_{t_{n-M+1}},S_{t_{n-M+2}},\dots,S_{t_{n}}$ are required to write
the law of the couple $\left(S_{t_{k}},A_{k-M+1}^{k}\right)$ for
any $k$ in $\left\{ n+1,\dots,n+M\right\} $. So, \citet{kao2003}
suggests to use the moving window process $\left(\mathbf{S}_{n}\right)_{n=M,\dots,N}$
defined as 

\begin{align*}
\mathbf{S}_{n} & =\left(\mathbf{S}_{n,1},\mathbf{S}_{n,2},\dots,\mathbf{S}_{n,M}\right)^{\top}=\left(S_{t_{n-M+1}},\dots,S_{t_{n}}\right)^{\top},
\end{align*}
which is Markovian, although process $\left(S_{t_{n}},A_{n-M+1}^{n}\right)$
is not, and $A_{n-M+1}^{n}$ is measurable with respect to $\mathbf{S}_{n}$. 

Here we prefer to consider a different approach. Specifically, we
describe the moving average option in terms of the process of the
last partial averages $\left(\mathbf{A}_{n}\right)_{n=M,\dots,N}$,
which writes
\begin{align*}
\mathbf{A}_{n} & =\left(\mathbf{A}_{n,1},\mathbf{A}_{n,2},\dots,\mathbf{A}_{n,d_{n}^{A}}\right){}^{\top}=\left(A_{n-M+1}^{n},A_{n-M+2}^{n},\dots,A_{\min\left\{ n-1,N-M+1\right\} }^{n},A_{n}^{n}\right)^{\top}.
\end{align*}
For example, if $M=2$, then $\mathbf{A}_{n}=\left(A_{n-1}^{n},\dots,A_{\min\left\{ n-1,N-1\right\} }^{n},A_{n}^{n}\right)^{\top}$,
thus $\mathbf{A}_{N}=\left(A_{N-1}^{N},A_{N}^{N}\right)^{\top}$,
$\mathbf{A}_{N-1}=\left(A_{N-2}^{N-1},A_{N-1}^{N-1}\right)^{\top}$
and $\mathbf{A}_{N-2}=\left(A_{N-3}^{N-2},A_{N-2}^{N-2}\right)^{\top}$,
while if $M=3$, then $\mathbf{A}_{n}=\left(A_{n-2}^{n},\dots,A_{\min\left\{ n-1,N-2\right\} }^{n},A_{n}^{n}\right)^{\top}$,
thus $\mathbf{A}_{N}=\left(A_{N-2}^{N},A_{N}^{N}\right)^{\top}$,
$\mathbf{A}_{N-1}=\left(A_{N-3}^{N-1},A_{N-2}^{N-1},A_{N-1}^{N-1}\right)^{\top}$
and $\mathbf{A}_{N-2}=\left(A_{N-4}^{N-2},A_{N-3}^{N-2},A_{N-2}^{N-2}\right)^{\top}$.

Let us point out some properties of the process $\mathbf{A}$. The
dimension changes with time and, at time $t_{n}$, it is equal to
$d_{n}^{A}=\min\left\{ M,N-n+2\right\} $ which means it increases
by one unit moving backward in time up to $t_{N-M+2}$ when it is
equal to $M$ and stops augmenting. Specifically, the last component
is $A_{n}^{n}=S_{t_{n}}$, while the first component is the average
of the last $M$ observed underlying values, the second one is the
average of the last $M-1$ observed underlying values and so on. Generally
speaking, the $i$-th component $\mathbf{A}_{n,i}$ is equal to the
mean of the last $M+1-i$ observed underlying values, with the exception
of the last component $\mathbf{A}_{n,d_{n}}$ which is the last observed
price, that is $S_{n}$. If $S$ is Markovian, then $\mathbf{A}$
is Markovian too. In fact, $A_{n}^{n}=S_{t_{n}}$ and
\begin{align}
A_{n+1}^{n+1} & =S_{t_{n+1}},\\
A_{i}^{n+1} & =\frac{\left(n-i+1\right)A_{i}^{n}+S_{t_{n+1}}}{n-i+2},
\end{align}
or equivalently, with respect to the components of $\mathbf{A}$,
we have
\begin{equation}
\mathbf{A}_{n+1,i}=\begin{cases}
S_{t_{n+1}} & \text{if }i=d_{n+1}^{A}\\
\frac{\left(M-i\right)\mathbf{A}_{n,i+1}+S_{t_{n+1}}}{M+1-i} & \text{otherwise. }
\end{cases}\label{eq:anpu}
\end{equation}

We emphasize that the payoff of the option is measurable with respect
to process $\mathbf{A}$, so let us denote with $\Psi_{n}^{\mathbf{A}}$
the payoff as a function of the process $\mathbf{A}_{n}$, that is
\begin{equation}
\Psi_{n}^{\mathbf{A}}\left(A_{n}\right)=\Psi_{n}^{\mathbf{A}}\left(A_{n-M+1}^{n},A_{n-M+2}^{n},\dots,A_{\min\left\{ n-1,N-M+1\right\} }^{n},A_{n}^{n}\right)=\max\left(0,A_{n}^{n}-A_{n-M+1}^{n}\right).
\end{equation}
The use of the process $\mathbf{A}$ in place of the process $\mathbf{S}$
brings computational advantages since, while process $\mathbf{S}$
is $M$-dimension, process $\mathbf{A}$ has dimension at most $M$,
so it speeds up calculations, in particular regression, which is a
key part in the GPR-GHQ method.

\section{The GRP-GHQ method }

The method we propose is called GPR-GHQ as it employs the Gaussian
Process Regression (GPR) method and the Gauss-Hermite (GHQ) quadrature
scheme. 

GPR is a non-parametric Bayesian method for regression that belongs
to the group of machine learning techniques. In the  \ref{sec:Gaussian-Process-Regression}
we report a brief description of such an approach. We use this technique
as GPR has several advantages over other similar techniques such as
neural networks: it works well on small data sets and has the ability
to provide uncertainty measurements on predictions. 

Finally, GHQ is a famous technique in option pricing, so we simply
present a brief description of it in \ref{sec:Gauss-Hermite-quadrature}.

\subsection{The algorithm for the Black-Scholes model}

The GPR-GHQ method is a backward induction algorithm that employs
GHQ to compute the continuation value of the option only for some
particular path of the underlying and it employs GPR to extrapolate
the whole continuation value from those observations. First of all,
we present the method for the Black-Scholes model and then we discuss
how to adapt it to the Clewlow-Strickland model and to the rough-Bergomi
model. So, let us suppose that the underlying follows the Black-Scholes
dynamics.

Let $N$ denote the number of time steps, $\Delta t=T/N$ be the time
increment and $t_{n}=n\,\Delta t$ represent the discrete time steps
for $n=0,1,\ldots,N$ as in Section \ref{sec:Moving-Average-options}.
In order to price the Bermudan moving average option, we consider
the process $\left(\mathbf{B}_{n}\right)_{n=M,\dots,N}$, which writes
\begin{align}
\mathbf{B}_{n} & =\left(\mathbf{B}_{n,1},\mathbf{B}_{n,2},\dots,\mathbf{B}_{n,d_{n}^{B}}\right){}^{\top}=\left(A_{n-M+2}^{n},A_{n-M+3}^{n},\dots,A_{\min\left\{ n-1,N-M+1\right\} }^{n},A_{n}^{n}\right)^{\top},\label{eq:B}
\end{align}
which can be obtained by $\mathbf{A}_{n}$ by dropping the first component
$\mathbf{A}_{n,1}$, that is $A_{n-M+1}^{n}$ (thus, the dimension
of $\mathbf{B}_{n}$ is $d_{n}^{B}=d_{n}^{A}-1)$. We compute the
option price by moving backward in time and by computing the option
value only at time steps $t_{n}$ with $n\in\left\{ M,\dots,N\right\} $.
Specifically, the option value $\mathcal{V}_{n}$ at time $t_{n}$
is determined by the process of partial averages $\mathbf{A}_{n}$
as follows:
\begin{equation}
\mathcal{V}_{n}\left(\mathbf{A}_{n}\right)=\max\left(\Psi_{n}^{\mathbf{A}}\left(\mathbf{A}_{n}\right),\mathcal{C}_{n}\left(\mathbf{B}_{n}\right)\right),\label{eq:update}
\end{equation}
 with $\mathcal{C}_{n}$ the continuation value function of the moving
average option at time $t_{n}$ which is given by the following relation
\begin{equation}
\mathcal{C}_{n}\left(\mathbf{B}_{n}\right)=\mathbb{E}_{t_{n},\mathbf{B}_{n}}\left[e^{-r\Delta t}\mathcal{V}_{n+1}\left(\mathbf{A}_{n+1}\right)\right].\label{eq:CV}
\end{equation}
where $\mathbb{E}_{t_{n},\mathbf{B}_{n}}$ represents the expectation
at time $t_{n}$ given that $\mathbf{B}_{n}$ is the value of the
process $\mathbf{B}$ at time $t_{n}$. Please observe that the law
of $\mathbf{A}_{n+1}|\mathbf{A}_{n}$ is the same as $\mathbf{A}_{n+1}|\mathbf{B}_{n}$
because, according to (\ref{eq:anpu}), $\mathbf{A}_{n,1}$ is not
used to obtain $\mathbf{A}_{n+1}$, so we can consider the conditional
expectation $\mathbb{E}_{t_{n},\mathbf{B}_{n}}$ that requires less
information with respect to $\mathbb{E}_{t_{n},\mathbf{A}_{n}}$ (so
it is more convenient from a computational perspective) and we write
$\mathcal{C}_{n}$ as a function of $\mathbf{B}_{n}$ in place of
$\mathbf{A}_{n}$ . Moreover, the following equation points out the
relation between $\mathbf{A}_{n+1}$ and $\mathbf{B}_{n+1}$
\begin{equation}
\mathbf{A}_{n+1}=\left(\frac{\left(M-1\right)\mathbf{B}_{n,1}+\mathbf{B}_{n+1,d_{n+1}^{B}}}{M},\mathbf{B}_{n+1}\right),\label{eq:ABnp1}
\end{equation}
so that one can obtain $\mathbf{A}_{n+1}$ from $\mathbf{B}_{n}$
and $\mathbf{B}_{n+1}$.

We stress out that the continuation value at maturity is zero, since
the option expiry after time $T$, that is
\begin{eqnarray}
\mathcal{C}_{N}\left(\mathbf{B}_{N}\right) & = & 0\label{eq:C0}
\end{eqnarray}
an obviously 
\[
\mathcal{V}_{N}\left(\mathbf{A}_{N}\right)=\Psi_{N}^{\mathbf{A}}\left(\mathbf{A}_{N}\right).
\]
Thus, we can observe that the continuation value at time $t_{N-1}$
simplifies as follows 
\begin{align}
\mathcal{C}_{N-1}\left(\mathbf{B}_{N-1}\right) & =\mathbb{E}_{t_{N-1},\mathbf{B}_{N-1}}\left[e^{-r\Delta t}\Psi_{N}^{\mathbf{A}}\left(\mathbf{A}_{N}\right)\right]\nonumber \\
 & =\mathbb{E}_{t_{N-1},\mathbf{B}_{N-1}}\left[e^{-r\Delta t}\max\left(0,S_{T}-\frac{1}{M}\sum_{j=N-M+1}^{N}S_{t_{j}}\right)\right]\nonumber \\
 & =\mathbb{E}_{t_{N-1},\mathbf{B}_{N-1}}\left[e^{-r\Delta t}\max\left(0,\frac{M-1}{M}S_{T}-\frac{1}{M}\sum_{j=N-M+1}^{N-1}S_{t_{j}}\right)\right]\nonumber \\
 & =\frac{M-1}{M}\mathbb{E}_{t_{N-1},\mathbf{B}_{N-1}}\left[e^{-r\Delta t}\max\left(0,S_{T}-\frac{1}{M-1}\sum_{j=N-M+1}^{N-1}S_{t_{j}}\right)\right]\nonumber \\
 & =\frac{M-1}{M}\mathbb{E}_{t_{N-1},\mathbf{B}_{N-1}}\left[e^{-r\Delta t}\max\left(0,S_{T}-\frac{MA_{N-1-M+2}^{N-1}}{M-1}\right)\right]\nonumber \\
 & =\frac{M-1}{M}\mathcal{C}all\left(t_{N-1},t_{N},A_{N-1}^{N-1},\frac{MA_{N-1-M+2}^{N-1}}{M-1}\right),\label{eq:CCAll}
\end{align}
where $\mathcal{C}all\left(t_{0},T,S_{0},K\right)$ stands for the
price of a European call option on $S$ with inception time $t_{0}$,
maturity $T$, spot value $S_{0}$, and strike $K$.

By exploiting equations (\ref{eq:C0}), (\ref{eq:CCAll}) and (\ref{eq:update}),
we can write a dynamic programming problem of the function $C$ as
follows:

\begin{equation}
\begin{cases}
\mathcal{C}_{N}\left(\mathbf{B}_{N}\right) & =0\\
\mathcal{C}_{N-1}\left(\mathbf{B}_{N-1}\right) & =\mathcal{C}all\left(t_{N-1},t_{N},A_{N-1}^{N-1},\frac{MA_{n-M+1}^{n}-A_{n}^{n}}{M-1}\right)\\
\mathcal{C}_{n}\left(\mathbf{B}_{n}\right) & =\mathbb{E}_{t_{n},\mathbf{B}_{n}}\left[e^{-r\Delta t}\max\left(\Psi_{n+1}^{A}\left(\mathbf{A}_{n+1}\right),\mathcal{C}_{n+1}\left(\mathbf{B}_{n+1}\right)\right)\right],\ \text{for}\ n=N-2,\dots,M.
\end{cases}\label{eq:BIA}
\end{equation}
So, by simply computing the price of a European Call option, the continuation
value can be evaluated at time $t_{N}$ and $t_{N-1}$, 

Now, we aim to solve problem (\ref{eq:BIA}) by moving backward in
time, starting from $t_{N-1}$ up to $t_{M}$. The key point is that,
for $n=N-2,\dots,M$, the continuation value is the expectation of
the random variable
\[
e^{-r\Delta t}\max\left(\Psi_{n+1}^{A}\left(\mathbf{A}_{n+1}\right),\mathcal{C}_{n+1}\left(\mathbf{B}_{n+1}\right)\right)|\mathbf{B}_{n},
\]
 which is measurable once the value of the one-dimensional random
variable $S_{t_{n+1}}$ is known. Therefore, if one can evaluate the
function $\mathcal{C}_{n+1}$, then $\mathcal{C}_{n}$ can be computed
efficiently at any point by means of a one-dimensional quadrature
formula which is a very efficient method. By iterating the same reasoning,
if the function $\mathcal{C}_{n}$ can be computed at any point, one
can also compute $\mathcal{C}_{n-1}$ and the procedure can repeat
up to $t_{M}$. However, if the evaluation of $\mathcal{C}_{n}$ at
any point exploits directly a quadrature formula, then the evaluation
of $\mathcal{C}_{n-1}$ , which depends on $\mathcal{C}_{n}$, would
require a two nested quadrature formulas. Similarly, evaluating $\mathcal{C}_{n-2}$
would require three nested quadrature formulas, so that the computational
cost explodes in a few time steps. In order to overcome such a problem,
we exploit GPR to learn $\mathcal{C}_{n}$ from a few observed value,
so that evaluating $\mathcal{C}_{n}$ at any point becomes very fast
as it relies directly on GPR.

First of all, we need to select some points where evaluate the function
$\mathcal{C}_{n}$ and employ those observations as the training set
for GPR. To this aim, we employ a quasi-random generator, for example
the Halton sequence, to simulate $P$ discrete time paths for the
process $S$ and thus for $\mathbf{A}$ and $\mathbf{B}$. 

Let us consider a time step $t_{n}$ and suppose that the continuation
value $\mathcal{C}_{n}$ is known, at least in an approximate form,
at time $t_{n+1}$. Our target is to obtain an approximation for $\mathcal{C}_{n}$.
To this aim, we consider a set $X^{n}$ of $P$ points whose elements
are the simulated values for $\mathbf{B}_{n}$:
\begin{equation}
X^{n}=\left\{ \mathbf{x}^{n,p}=\left(x_{1}^{n,p},\dots,x_{d_{n}^{B}}^{n,p}\right),p=1,\dots,P\right\} \subset\mathbb{R}^{d_{n}}.\label{eq:X}
\end{equation}
The GPR-GH method assesses $\mathcal{C}_{n}\left(\mathbf{x}^{n,p}\right)$
for each $\mathbf{x}^{n,p}\in X^{n}$ through $Q$-points GHQ. Specifically,
let $\left\{ u_{q}\right\} _{q=1,\dots,Q}$ and $\left\{ w_{q}\right\} _{q=1,\dots,Q}$
be the GHQ nodes and weights. The last component of $\mathbf{x}^{n,p}$,
i.e. $x_{d_{n}^{B}}^{n,p}$, corresponds to the value $S_{t_{n}}$,
so $Q$ possible determinations for $S_{t_{n+1}}|S_{t_{n}}=x_{d_{n}^{B}}^{n,p}$
are employed by GHQ, precisely,
\[
S^{n,p,q}=x_{d_{n}}^{n,p}\cdot\exp\left(\left(r-\frac{\sigma^{2}}{2}\right)\Delta t+u_{q}\sqrt{2\Delta t}\sigma\right),\ q=1,\dots,Q.
\]
So, there also are $Q$ possible determinations for $\mathbf{B}_{n+1}|\mathbf{B}_{n}=\mathbf{x}^{n,p}$,
specifically
\[
\tilde{X}^{n,p}=\left\{ \mathbf{\mathbf{\tilde{x}}}^{n,p,q}=\left(\tilde{x}_{1}^{n,p,q},\dots,\tilde{x}_{d_{n+1}^{B}}^{n,p,q}\right)\right\} _{q=1\dots Q},
\]
with 
\[
\tilde{x}_{d_{n+1}}^{n,p,q}=S^{n,p,q},
\]
and by exploiting relation (\ref{eq:anpu}), we can write
\[
\tilde{x}_{i}^{n,p,q}=\frac{\left(M-i-1\right)x_{i+1}^{n,q}+S^{n,p,q}}{M-i}.
\]
We also define 
\[
\tilde{x}_{0}^{n,p,q}=\frac{\left(M-1\right)x_{1}^{n,q}+S^{n,p,q}}{M},
\]
so that the vector $\left(\tilde{x}_{0}^{n,p,q},\mathbf{\mathbf{\tilde{x}}}^{n,p,q}\right)$
is a possible outcome for $\mathbf{A}_{n+1}|\mathbf{B}_{n}=\mathbf{x}^{n,p}$,
thanks to (\ref{eq:ABnp1}). Then, the continuation value at $\mathbf{x}^{n,p}$,
that is
\[
\mathcal{C}_{n}\left(\mathbf{x}^{n,p}\right)=\mathbb{E}_{t_{n},\mathbf{x}^{n,p}}\left[e^{-r\Delta t}\max\left(\Psi_{n+1}^{A}\left(\mathbf{A}_{n+1}\right),\mathcal{C}_{n+1}\left(\mathbf{B}_{n+1}\right)\right)\right],
\]
 is approximated by 
\begin{equation}
\mathcal{C}_{n}^{GHQ}\left(\mathbf{x}^{n,p}\right)=e^{-r\Delta t}\sum_{q=1}^{Q}w_{q}\max\left(\Psi_{n+1}^{A}\left(\tilde{x}_{0}^{n,p,q},\mathbf{\mathbf{\tilde{x}}}^{n,p,q}\right),\mathcal{C}_{n+1}\left(\mathbf{\mathbf{\tilde{x}}}^{n,p,q}\right)\right).\label{eq:update2}
\end{equation}
Equation (\ref{eq:update2}) can be evaluated only if the quantities
$\mathcal{C}_{n+1}\left(\mathbf{\mathbf{\tilde{x}}}^{n,p,q}\right)$
are known for all the future points $\mathbf{\mathbf{\tilde{x}}}^{n,p,q}$.
As pointed out in (\ref{eq:BIA}), both the functions $\mathcal{C}_{N}$
and $\mathcal{C}_{N-1}$ are known, so one can employ (\ref{eq:update2})
to compute $\mathcal{C}_{N-2}^{GHQ}\left(\mathbf{x}^{N-2,p}\right)$
for all $\mathbf{x}^{N-2,p}$ in $X^{N-2}$. In order to compute $\mathcal{C}_{N-3}\left(\mathbf{x}^{N-3,p}\right)$
for all $\mathbf{x}^{N-3,p}\in X^{N-3}$, and thus going on up to
$t_{M}$, the function $\mathcal{C}_{N-1}$ needs to be evaluated
for all the points in $\tilde{X}^{N-3}=\bigcup_{p=1}^{P}\tilde{X}^{N-3,p}$,
but we only know $\mathcal{C}_{N-2}^{GHQ}$ at $X^{N-2}$. To overcome
this issue, we employ the GPR method to approximate the function $\mathcal{C}_{N-2}$
at any point of $\mathbb{R}^{d}$ and in particular at the elements
of $\tilde{X}^{N-3}$. Specifically, let $\mathcal{C}_{N-2}^{GPR}$
denote the GPR prediction of $\mathcal{C}_{N-2}^{GHQ}$, obtained
by considering the predictor set $X^{N-2}$ and the response $\mathbf{y}\in\mathbb{R}^{P}$
given by 
\begin{equation}
y^{p}=\mathcal{C}_{N-2}^{GHQ}\left(\mathbf{x}^{n,p}\right),\ p\in\left\{ 1,\dots,P\right\} .
\end{equation}
The GPR-GHQ approximation $\mathcal{C}_{N-2}^{GPR-GHQ}$ of the value
function $\mathcal{C}_{N-3}$ at time $t_{N-3}$ can be computed as
follows:
\[
\mathcal{C}_{N-3}^{GPR-GHQ}\left(\mathbf{x}^{N-3,p}\right)=e^{-r\Delta t}\sum_{q=1}^{Q}w_{q}\max\left(\Psi_{N-2}^{A}\left(\tilde{x}_{0}^{N-3,p,q},\mathbf{\mathbf{\tilde{x}}}^{N-3,p,q}\right),\mathcal{C}_{N-2}^{GPR}\left(\mathbf{\mathbf{\tilde{x}}}^{N-3,p,q}\right)\right),\ p\in\left\{ 1,\dots,P\right\} .
\]
 The procedure described above for $n=N-3$ can be replicated for
any value $n$ from $N-3$ to $M$, so that the dynamic programming
problem can be solved. Specifically, let $n\in\left\{ M,\dots,N-4\right\} $
and let $\mathcal{C}_{n+1}^{GPR}$ denote the GPR prediction of $\mathcal{C}_{n+1}^{GPR-GHQ}$
obtained from predictor set $X^{n+1}$ and the response $\mathbf{y}\in\mathbb{R}^{P}$
given by 
\begin{equation}
y^{p}=\mathcal{C}_{n+1}^{GPR-GHQ}\left(\mathbf{x}^{p}\right).
\end{equation}
Then, the function $\mathcal{C}_{n}^{GPR-GHQ}$ is defined as follows:
\[
\mathcal{C}_{n}^{GPR-GHQ}\left(\mathbf{x}^{n,p}\right)=e^{-r\Delta t}\sum_{q=1}^{Q}w_{q}\max\left(\Psi_{n+1}^{A}\left(\tilde{x}_{0}^{n,p,q},\mathbf{\mathbf{\tilde{x}}}^{n,p,q}\right),\mathcal{C}_{n+1}^{GPR}\left(\mathbf{\mathbf{\tilde{x}}}^{n,p,q}\right)\right),\ p\in\left\{ 1,\dots,P\right\} 
\]
Once the function $\mathcal{C}_{M}^{GPR-GHQ}$ has been estimated,
the option price $\mathcal{V}_{0}$ at inception can be computed by
discounting the expected option value at time $t_{M}$, which is the
first time step the option can be exercised, that is
\begin{equation}
\mathcal{V}_{0}=e^{-rM\Delta t}\mathbb{E}\left[\max\left(\Psi_{M}^{A}\left(\mathbf{A}_{M}\right),\mathcal{C}_{M}^{GPR-GHQ}\left(\mathbf{B}_{M}\right)\right)\right].\label{eq:last}
\end{equation}
Finally, the expectation in (\ref{eq:last}) is computed by means
of a Monte Carlo approach with antithetic variates. 

\subsubsection{Similarity reduction\label{subsec:Similarity-reduction}}

The continuation value has an interesting scale property, explained
in the following Proposition.
\begin{prop}
\label{prop:1}Let $t_{n}$ be a step time with $n$ in $\left\{ M,\dots,N\right\} $.
Then, the continuation value for a moving average option in the Black-Scholes
model is positively homogeneous, that is for every positive real number
$\kappa$
\[
\mathcal{C}_{n}\left(\mathbf{B}\right)=\kappa\mathcal{C}_{n}\left(\frac{1}{\kappa}\mathbf{B}_{n}\right).
\]
\end{prop}

A proof of proposition \ref{prop:1} is presented in the  (\ref{sec:Proof-of-Proposition}). 

Now, if we set $\kappa=\mathbf{B}_{n,d_{n}^{B}}$ (that is $\kappa=S_{t_{n}}$,
which is strictly positive), we obtain
\[
\mathcal{C}_{n}\left(\mathbf{B}_{n}\right)=S_{t_{n}}\mathcal{C}_{n}\left(\frac{1}{S_{t_{n}}}\left(\mathbf{B}_{n,1},\dots,\mathbf{B}_{n,d_{n}^{B}-1}\right),1\right).
\]
Please observe that the vector $\frac{1}{S_{t_{n}}}\mathbf{B}_{n}$
has the last component equal to $1$, which can be dropped. Therefore,
in order to keep advantage of such a property, we define the process

\begin{align}
\mathbf{C}_{n} & =\left(\mathbf{C}_{n,1},\mathbf{C}_{n,2},\dots,\mathbf{C}_{n,d_{n}^{C}}\right){}^{\top}=\frac{1}{S_{t_{n}}}\left(A_{n-M+2}^{n},A_{n-M+3}^{n},\dots,A_{\min\left\{ n-1,N-M-1\right\} }^{n}\right)^{\top},\label{eq:C}
\end{align}
 that is the process $\mathbf{B}_{n}$ with its last component dropped
and the others divided by $S_{t_{n}}$, so the dimension of $\mathbf{C}_{n}$
is $d_{n}^{C}=d_{n}^{B}-1.$ Therefore, we can define the function
$\mathcal{C}^{SR}$ which represents the continuation value by assuming
the actual value of $S$ equal to $1$ and that satisfies the following
relation
\[
\mathcal{C}_{n}^{SR}\left(\mathbf{C}_{n}\right)=\frac{1}{S_{t_{n}}}\mathcal{C}_{n}\left(\mathbf{B}_{n}\right),
\]
or equivalently
\[
\mathcal{C}_{n}\left(\mathbf{B}_{n}\right)=\mathbf{B}_{n,d_{n}^{B}}C_{n}^{SR}\left(\mathbf{C}_{n}\right).
\]
Thus, at any time $t_{n}$, the continuation value is given through
the function $\mathcal{C}_{n}^{SR}$ in place instead of function
$\mathcal{C}_{n}$. We stress out that function $\mathcal{C}_{n}^{SR}$
has one variable less than $\mathcal{C}_{n}$ so it is easier to be
learned by the GPR method. Moreover, if $M=2$, the number of variables
of $\mathcal{C}_{n}^{SR}$ is zero, that is the continuation value
is fully described by a number, that is the continuation value for
$S_{t_{n}}=1$. In this particular case, the use of GPR can be avoided.

\subsection{Adaptations for the Clewlow-Strickland model }

As far as the Clewlow-Strickland model is considered in place of the
Black-Scholes models, the main difference is related to the use of
the GHQ. Let us define $Y_{t}=\ln\left(S_{t}\right)-\beta\left(t\right)$,
with
\begin{equation}
\beta\left(t\right)=\ln\left(F(0,t)\right)+\frac{\sigma^{2}}{4\alpha}\left(e^{-2\alpha t}-1\right).\label{eq:CS_beta}
\end{equation}
Then, for all $s<t$, the random variable $Y_{t}|Y_{s}$ has a normal
distribution. In particular: 
\begin{equation}
Y_{t}|Y_{s}\sim\mathcal{N}\left(e^{-\alpha\left(t-s\right)}Y_{s},\frac{\sigma^{2}}{2\alpha}\left(1-e^{-2\alpha(t-s)}\right)\right).\label{eq:CS_Y}
\end{equation}
By exploiting formulas (\ref{eq:CS_beta}) and (\ref{eq:CS_Y}) one
can easily simulate the path of the process $\left(Y_{t}\right)_{t\in\left[0,T\right]}$
and thus of $\left(S_{t}\right)_{t\in\left[0,T\right]}$ . In particular

\begin{eqnarray}
S_{t_{n+1}} & = & \exp\left(e^{-\alpha\Delta t}\left(\ln\left(S_{t_{n}}\right)-\beta\left(t_{n}\right)\right)+\sqrt{\frac{\sigma^{2}}{2\alpha}\left(1-e^{-2\alpha(t-s)}\right)}G+\beta\left(t_{n+1}\right)\right),\label{eq:sim_CS}
\end{eqnarray}
with $G\sim\mathcal{N}\left(0,1\right).$ Now, let $\left\{ u_{q}\right\} _{q=1,\dots,Q}$
and $\left\{ w_{q}\right\} _{q=1,\dots,Q}$ be the Gauss-Hermite quadrature
nodes and weights. Let $X^{n}$ be as defined in (\ref{eq:X}) and
let us consider a point $\mathbf{x}^{n,p}=\left(x_{1}^{n,p},\dots,x_{d_{n}^{B}}^{n,p}\right)$
of $X^{n}$, which represents a value of $\mathbf{B}_{n}$. Then,
according to (\ref{eq:sim_CS}), GHQ considers $Q$ possible determinations
for $S_{t_{n+1}}|S_{t_{n}}=x_{d_{n}^{B}}^{n,p}$, given by
\[
S^{n,p,q}=\exp\left(e^{-\alpha\Delta t}\left(\ln\left(x_{d_{n}}^{n,p}\right)-\beta\left(t_{n}\right)\right)+\sqrt{\frac{\sigma^{2}}{2\alpha}\left(1-e^{-2\alpha(t-s)}\right)}u_{q}+\beta\left(t_{n+1}\right)\right).
\]

\subsection{Adaptations for the rough-Bergomi model}

As far as the rough-Bergomi is considered, the simulation of the quasi-random
paths follows the scheme presented in  (\ref{sec:Simulation-of-Stochastic}).
This model is two-dimensional as both the volatility $V$ and the
underlying price $S$ are stochastic. Moreover, since volatility is
a non Markovian process, the continuation value at time $t_{n}$ should
depend on all the passed values of $\left(S,V\right)$, that is $\left\{ \left(V_{t},S_{t}\right):0\leq t\leq t_{n}\right\} $,
which leads to an infinite-dimensional problem. A first approximation
is obtained by replacing the time continuous process $\left(V,S\right)_{t\in\left[0,T\right]}$
with the discrete time process obtained from the Euler-Maruyama scheme
(\ref{eq:EM}). For sake of simplicity, we denote with $\left(V_{t_{n}},S_{t_{n}}\right)_{n=1,\dots N}$
such a discrete time process. In particular, (see  (\ref{sec:Simulation-of-Stochastic}))
\[
V_{t_{n}}=\xi_{0}\exp\left(-\frac{1}{2}\eta^{2}\left(t_{n}\right)^{2H}+\eta\widetilde{W}_{t_{n}}^{H}\right),
\]
with $\widetilde{W}_{t_{n}}^{H}$ obtained by the following scalar
product 
\[
\widetilde{W}_{t_{n}}^{H}=\Lambda_{2n,1:2n}\left(G_{1},\dots,G_{2n}\right)^{\top},
\]
where $\Lambda_{2n,1:2n}$ stands for the elements of the $2n-th$
row of $\Lambda$, in the column position from $1$ to $2n$. 

In order to solve the control problem, we consider the filtration
$\mathcal{F}_{n}$ generated by the $2n$ variables $W_{t_{1}}^{1},\widetilde{W}_{t_{1}}^{H},\dots,$
$W_{t_{n}}^{1},\widetilde{W}_{t_{n}}^{H}$, which is equivalent to
the filtration generated by $S_{t_{1}},V_{t_{1}},\dots,S_{t_{n}},V_{t_{n}}$.
Let $n\geq k$. We can write 
\[
V_{t_{n}}=\xi_{0}\exp\left(-\frac{1}{2}\eta^{2}\left(t_{n}\right)^{2H}+\eta\left[\Lambda_{2j,1:2k}\left(G_{1},\dots,G_{2k}\right)^{\top}+\Lambda_{2j,\left(2k+1\right):2j}\left(G_{2k+1},\dots,G_{2j}\right)^{\top}\right]\right),
\]
being $\left(G_{2k+1},\dots,G_{2j}\right)$ independent form $\mathcal{F}_{k}$.
Moreover, if $\Lambda_{1:2k,1:2k}$ stands for the squared matrix
obtained by selecting the first $2k$ rows and $2k$ columns of $\Lambda,$
then 
\[
\left(G_{1},\dots,G_{2k}\right)^{\top}=\left(\Lambda_{1:2k,1:2k}\right)^{-1}\left(\Delta W_{1}^{1},\widetilde{W}_{t_{1}}^{H},\dots,\Delta W_{k}^{1},\widetilde{W}_{t_{k}}^{H}\right)^{\top},
\]
 which implies that $\left(G_{1},\dots,G_{2k}\right)^{\top}$ is measurable
with respect to $\mathcal{F}_{k}$. Thus if $n\geq k$ then the random
variables $V_{t_{n}}\left|\mathcal{F}_{k}\right.$ and $V_{t_{n}}\left|E_{n,k}\right.$have
the same law, being
\begin{equation}
E_{n,k}=\Lambda_{2n,1:2k}\left(G_{1},\dots,G_{2k}\right)^{\top}.\label{eq:E}
\end{equation}

The continuation value at time $t_{n}$ depends on the values $\left\{ E_{h,n}\right\} _{h=n,\dots M,}$
as all these values impact on the law of the future volatility. These
values should be all included in the list of the predictors of the
GPR, but their number may be large, so, inspired by \citet{bayer2020},
we consider a non-negative integer $J$ and include in the set of
predictors only the actual value of $E_{n,n}=\widetilde{W}_{t_{n}}^{H}$
and at most the next $J$ values, that is $\left\{ E_{h,n}\right\} _{h=n+1,\dots\min\left(N,n+J\right)}$,
so $d_{n}^{E}=\min\left(N-n,J\right)+1$ elements.

So, as far as the rough-Bergomi model is considered, the elements
of $X^{n}$ are vectors in $\mathbb{R}^{d_{n}^{E}+d_{n}^{B}}$. The
first $d_{n}^{E}$ components are quasi-random simulation of the variables
$\left\{ E_{h,n}\right\} _{h=n,\dots\min\left(M,n+J\right)}$ and
the next $d_{n}^{B}$ components are quasi-random simulation of the
process $\mathbf{B}_{n}$, both obtained from the initial quasi-random
simulations.

Finally we stress out that the computation of the continuation value
is performed through a bi-dimensional GHQ formula. Specifically, let
$\left\{ u_{q}\right\} _{q=1,\dots,Q}$ and $\left\{ w_{q}\right\} _{q=1,\dots,Q}$
be the Gauss-Hermite quadrature nodes and weights and let $\mathbf{x}^{n,p}$
be a point of $X^{n}$ and let $\mathbf{G}^{n,p}=\left(G_{1}^{p},\dots,G_{2n}^{p}\right)^{\top}$
be the quasi-random normal vector used to generate the variables $\left(\Delta W_{1}^{1},\widetilde{W}_{t_{1}}^{H},\dots,\Delta W_{n}^{1},\widetilde{W}_{t_{n}}^{H}\right)$
that have lead to $\mathbf{x}^{n,p}$ through (\ref{eq:EM}), (\ref{eq:B})
and (\ref{eq:E}). The elements of the set 
\[
\tilde{X}^{n,p}=\left\{ \mathbf{\mathbf{\tilde{x}}}^{n,p,q_{1},q_{2}}=\left(\tilde{x}_{1}^{n,p,q_{1},q_{2}},\dots,\tilde{x}_{d_{n+1}^{B}}^{n,p,q_{1},q_{2}}\right)\right\} _{q_{1},q_{2}=1\dots Q}
\]
are obtained by using (\ref{eq:EM}), (\ref{eq:B}) and (\ref{eq:E})
as for $\mathbf{x}^{n+1,p}$, but replacing $\mathbf{G}^{n+1,p}$
with 
\[
\mathbf{\tilde{G}}^{n,p,q_{1},q_{2}}=\left(G_{1}^{p},\dots,G_{2n}^{p},u_{q_{1}},u_{q_{2}}\right)^{\top}.
\]

\section{Binomial chain}

Similarity reduction introduced in Subsection (\ref{subsec:Similarity-reduction})
can be exploited to define an efficient approach based on a binomial
tree in the Black-Scholes model. 

The use of the CRR binomial tree for pricing moving average options
has already be investigated by \citet{kao2003} but his approach cannot
be applied for long averaging window since the required memory and
computational time grows exponentially with $M$. 

Before presenting our binomial approach, let us recall the main features
of the binomial method of Kao and Lyuu. The algorithm exploits a recombination
binomial tree with $N$ time steps and associates at each node $2^{M-1}$
vectors in $\left\{ 0,1\right\} ^{M-1}$ which represent the possible
moves in the tree that have lead to that node along the tree. In particular,
$0$ represents a down move and $1$ an up move, so that the total
number of possible states in the tree is $O\left(N^{2}2^{M}\right)$
and the computational cost is $O\left(N^{2}2^{M+1}\right)$. To be
precise, such an approach is then generalized by considering a $\left(L+1\right)$-nomial
tree for a certain positive integer $L$, but for sake of simplicity,
we limit our discussion to the binomial case. 

Following Kao, we consider a Markov chain defined on a CRR binomial
tree. The set of all possible states at time $t_{n}$ for $n\geq M$
is 
\[
\left\{ \left(\mathbf{s}_{p},S_{n,k}\right),p=1,\dots,2^{M-1},k=0,\dots,n,\right\} ,
\]
with 
\[
\mathbf{s}_{p}=\left(\mathbf{s}_{p,1},\dots,\mathbf{s}_{p,M-1}\right)^{\top},
\]
and 
\[
S_{n,k}=S_{0}e^{\left(2k-n\right)\sigma\sqrt{\Delta t}}.
\]
We stress out that the state $\left(\mathbf{s}_{p},S_{n,k}\right)$
encodes the value $\left(S_{t_{n-M+1}},\dots,S_{t_{n}}\right)$ with
\[
S_{t_{n-j}}=S_{n,k}\exp\left(-\sigma\sqrt{\Delta t}\sum_{i=M-j}^{M-1}\left(2\mathbf{s}_{p,i}-1\right)\right),\ j=0,\dots,M-1.
\]
Thus, the payoff for a state $\left(\mathbf{s}_{p},S_{n,k}\right)$
becomes
\[
\Psi^{CRR}\left(\mathbf{s}_{p},S_{n,k}\right)=\max\left(S_{n,k}-\frac{1}{M}\sum_{j=1}^{M}S_{n,k}\exp\left(-\sigma\sqrt{\Delta t}\sum_{i=M-j}^{M-1}\left(2\mathbf{s}_{p,i}-1\right)\right)\right).
\]

If the process state at time $t_{n}$ is $\left(\mathbf{s}_{p},S_{n,k}\right)$,
then the possible next states are denoted with $\left(\mathbf{s}_{p}^{\mathrm{up}},S_{n,k}^{\mathrm{up}}\right)$
and $\left(\mathbf{s}_{p}^{\mathrm{dw}},S_{n,k}^{\mathrm{dw}}\right)$.
In particular,
\[
\mathbf{s}_{p}^{\mathrm{up}}=\left(\mathbf{s}_{p,2},\dots,\mathbf{s}_{p,M-1},1\right),\ S_{n,k}^{\mathrm{up}}=S_{n,k}e^{\sigma\sqrt{\Delta t}}
\]
and 
\[
\mathbf{s}_{p}^{\mathrm{dw}}=\left(\mathbf{s}_{p,2},\dots,\mathbf{s}_{p,M-1},0\right),\ S_{n,k}^{\mathrm{dw}}=S_{n,k}e^{-\sigma\sqrt{\Delta t}}.
\]
Transition probabilities are 
\[
p^{\mathrm{up}}=\frac{\exp\left(r\Delta t\right)-\exp\left(-\sigma\Delta t\right)}{\exp\left(\sigma\Delta t\right)-\exp\left(-\sigma\Delta t\right)}\text{ and }p^{dw}=1-p^{\mathrm{up}}
\]
respectively. Option evaluation is performed by moving backward along
the tree. At maturity, continuation $\mathcal{C}_{N}^{CRR}$ value
is zero for all the states, that is 
\[
\mathcal{C}_{N}^{CRR}\left(\mathbf{s}_{p},S_{n,k}\right)=0\ \text{for all}\ p=1,\dots,2^{M-1}.
\]
The continuation value at a generic time step $n$ is defined discounting
the expectation of option value at next time step:

\begin{multline*}
\mathcal{C}_{n}^{CRR}\left(\mathbf{s}_{p},S_{n,k}\right)=e^{-r\Delta t}\left[p_{\mathrm{up}}\max\left(\Psi^{CRR}\left(\mathbf{s}_{p}^{\mathrm{up}},S_{n,k}^{\mathrm{up}}\right),\mathcal{C}_{n}^{CRR}\left(\mathbf{s}_{p}^{\mathrm{up}},S_{n,k}^{\mathrm{up}}\right)\right)\right.\\
\left.+p_{\mathrm{dw}}\max\left(\Psi^{CRR}\left(\mathbf{s}_{p}^{\mathrm{dw}},S_{n,k}^{\mathrm{dw}}\right),\mathcal{C}_{n}^{CRR}\left(\mathbf{s}_{p}^{\mathrm{dw}},S_{n,k}^{\mathrm{dw}}\right)\right)\right].
\end{multline*}
By exploiting backward induction, it is straightforward to prove that
$\mathcal{C}_{n}^{CRR}$ is positive homogeneous, that is 
\[
\mathcal{C}_{n}^{CRR}\left(\mathbf{s}_{p},S_{n,k}\right)=\kappa\mathcal{C}_{n}^{CRR}\left(\mathbf{s}_{p},\frac{1}{\kappa}S_{n,k}\right),
\]
for every $\kappa>0$ so we can drop the dependence by the specific
value of $S$. 

Similarity reduction allows us to improve the algorithm proposed by
\citet{kao2003}. The proposed binomial chain algorithm (BC) exploits
positive homogeneity to reduce memory consumption and computational
cost. Specifically, if we assume that the value of the underlying
associated to a node is always equal to $1$, the binomial tree with
$O\left(N^{2}\right)$ nodes collapses to a binomial Markov chain
with $N$ time steps, each of them consisting of $2^{M-1}$ states
$\left\{ \mathbf{s}_{p}\right\} _{p=1,\dots,2^{M-1}}=\left\{ 0,1\right\} ^{M-1}$
which again represent the possible past moves. Specifically, the next
states of $\mathbf{s}_{p}$ are again $\mathbf{s}_{p}^{\mathrm{up}}$
and $\mathbf{s}_{p}^{\mathrm{dw}}$ with probability $p^{\mathrm{up}}$
and $p^{dw}$ respectively. We can define a simplified continuation
value function $\mathcal{C}_{n}^{BC}$ by setting
\[
\mathcal{C}_{n}^{BC}\left(\mathbf{s}_{p}\right)=\mathcal{C}_{n}^{CRR}\left(\mathbf{s}_{p},1\right)
\]
and by exploiting similarity reduction, the recursive formula becomes 

{\small{}
\[
\begin{cases}
\mathcal{C}_{N}^{BC}\left(\mathbf{s}_{p}\right) & =0\\
\mathcal{C}_{n}^{BC}\left(\mathbf{s}_{p}\right) & =e^{-r\Delta t}\left[p_{\mathrm{up}}e^{\sigma\sqrt{\Delta t}}\max\left(\Psi^{CRR}\left(\mathbf{s}_{p}^{\mathrm{up}},1\right),\mathcal{C}_{n+1}^{BC}\left(\mathbf{s}_{p}^{\mathrm{up}}\right)\right)+p_{\mathrm{dw}}e^{-\sigma\sqrt{\Delta t}}\max\left(\Psi^{CRR}\left(\mathbf{s}_{p}^{\mathrm{dw}},1\right),\mathcal{C}_{n+1}^{BC}\left(\mathbf{s}_{p}^{\mathrm{dw}}\right)\right)\right].
\end{cases}
\]
}{\small\par}

Finally, once the continuation value is available at time step $M$
the option value at inception is obtained by averaging the option
value at the various states of the binomial chain at time $t_{M}$,
that is{\small{}
\begin{equation}
\mathcal{V}_{0}^{BC}=S_{0}e^{-r\Delta tM}\left(e^{\sigma\sqrt{\Delta t}}p_{\mathrm{up}}+e^{-\sigma\sqrt{\Delta t}}p_{\mathrm{dw}}\right)\sum_{p=1}^{2^{M-1}}P\left(\mathbf{s}_{p}\right)\left(e^{\sigma\sqrt{\Delta t}}\right)^{2\sum_{i=1}^{M-1}\mathbf{s}_{p,i}-M-1}\max\left(\Psi^{CRR}\left(\mathbf{s}_{p},1\right),\mathcal{C}_{M}^{BC}\left(\mathbf{s}_{p}\right)\right),
\end{equation}
}with 
\begin{equation}
P\left(\mathbf{s}_{p}\right)=\left(p_{\mathrm{up}}\right)^{\sum_{i=1}^{M-1}\mathbf{s}_{p,i}}\left(p_{\mathrm{dw}}\right)^{M-1-\sum_{i=1}^{M-1}\mathbf{s}_{p,i}}.
\end{equation}

We stress out that the binomial chain method returns exactly the same
prices as the CRR method, but its efficient implementation based on
similarity reduction allows one to reduced both the memory and computational
costs by a factor $N$, thus BC can manage longer averaging windows.
Specifically, the total number of possible states is $O\left(N2^{M}\right)$
and the computational cost is $O\left(N2^{M+1}\right)$.

\begin{figure}
\begin{centering}
\includegraphics[width=0.5\textwidth]{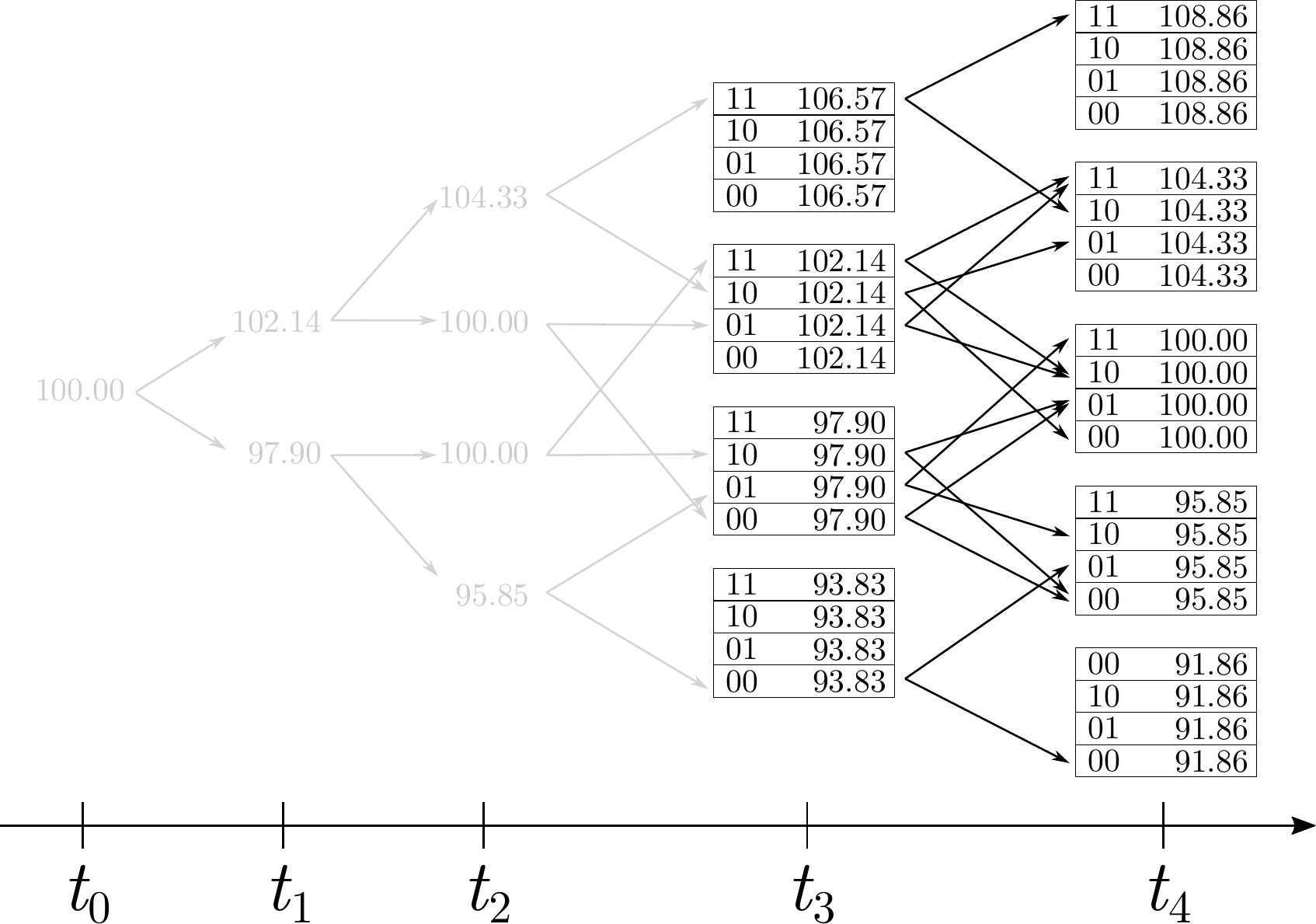}
\par\end{centering}
\vspace{10mm}
\begin{centering}
\includegraphics[width=0.5\textwidth]{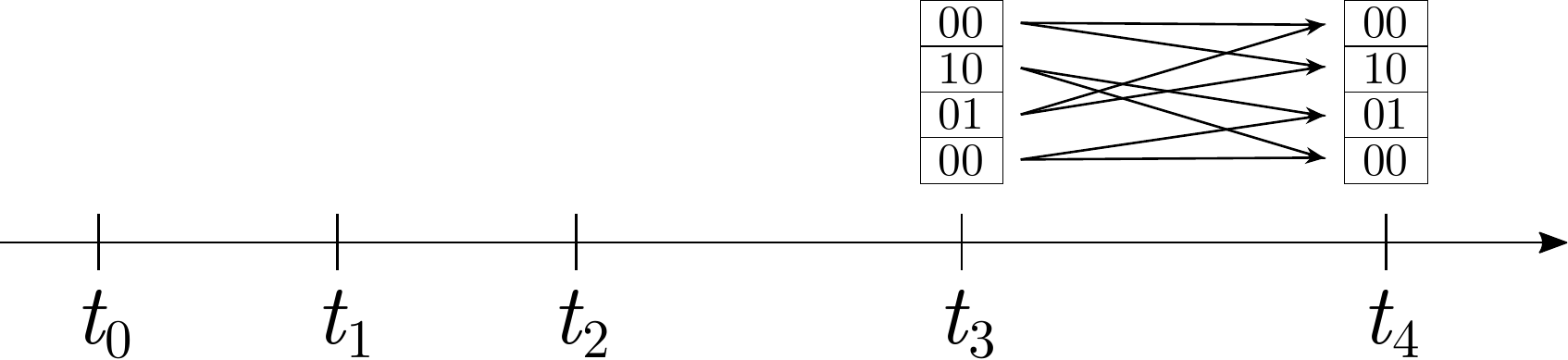}
\par\end{centering}
\caption{\label{fig:The-structure-of}The structure of the binomial CRR tree
and of the binomial chain for $N=4,M=3,r=0.05,\sigma=0.03$ and $T=0.02$.}

\end{figure}

Figure \ref{fig:The-structure-of} presents an example of the binomial
tree and of the binomial chain for $N=4$ and $M=3$. In both the
two figures, each rectangle contains a state of the system. As suggested
by this example, the number of possible states in the case of the
binomial chain is much lower than the number of states in the binomial
tree, which allows to reduce computational times.

\section{Numerical experiments}

In this Section we report the results of the numeric experiment. We
employ the GPR-GHQ method and we test it against a standard Longstaff-Schwartz
(LS) approach in all the three stochastic models. Specifically, we
report the results for both the two methods and we compare them against
the benchmark values. We also report the results by working with fixed
computing time, specifically, $30$ seconds, $1$ minute, and $2$
minutes. Finally, as far as the moving average option is considered,
we study two different maturities and time-step configurations, that
is $T=0.2$ with $N=50$ and $T=1.0$ with $N=250$, which correspond
to 50 or 250 trading days. We also present two benchmarks, which have
been computed through an independent forward Monte Carlo approach.
Specifically, the benchmark values are the average discounted payoff
values determined according to the optimal exercise strategy based
on the continuation value given by GPR-GHQ and by LS respectively.
The parameters used to compute the benchmarks are $deg=2$, $P=10^{7}$
for LS, and $Q=64$, $P=8000$ for GPR-GHQ. The number of Monte Carlo
paths for the forward step has been determined so that the radius
of the $95
$ confidence intervals is $0.01.$ For all prices obtained through
Monte Carlo procedures, we also report the margins of error at the
$95\%$ confidence level.

Numerical procedures have been implemented in MATLAB. The timed comparisons
were carried out on a personal computer equipped with an Intel i5-1035G1
processor and 8 GB of RAM, and only one core was used in order to
stabilize the calculation times as much as possible. In all other
cases, the numerical values were calculated using an 8-core Intel
Xeon Gold 6230 20C processor server with 16 GB RAM.

\subsection{The Black-Scholes model}

The parameters for the Black-Scholes model are the same used by \citet{bernhart2011}
(see Table (\ref{tab:BS})). 

Tables (\ref{tab:BS_2}), (\ref{tab:BS_2-1}) and (\ref{tab:BS_3})
report the results for LS, GPR-GHQ, and BC for different values of
$M$, by changing the numerical parameters of the methods. Specifically,
(\ref{tab:BS_2}) is referred to $T=0.2$, while Table (\ref{tab:BS_3})
is referred to $T=1.0$. Moreover, Table (\ref{tab:BS_2}) also reports
the results obtained by \citet{bernhart2011} and \citet{lelong2019}.
Finally, Table (\ref{tab:BS_4}) reports the results of the test with
fixed computational time. In this particular case, we exclude the
BC method from the comparison, as it does not allow to select the
discretization parameters in order to obtain a predetermined computational
time.

Looking at values in Table (\ref{tab:BS_2}), we can observe that
the results obtained with the three methods are substantially consistent
with each other and with the values in the literature. For small values
of $M$ (say $M\leq10$), LS and GPR-GHQ give very similar values,
while BC underestimates the exact value. For $M=20$, the three methods
return essentially the same price, however, BC should be preferred
as it stands out for its calculation speed, less than one second.
Also for $M=30$ the three methods provide similar prices, but the
results obtained through BC and GPR-GHQ are slightly higher than those
of LS, which indicates that LS loses effectiveness in case of large
size. This trend is confirmed by the comparison between the GPR-GHQ
and LS benchmarks: the continuation value provided by GPR-GHQ is more
accurate than that provided by LS, thus obtaining a better exercise
strategy. Note that the ability to approximate the continuation value
of LS for large values of $M$ is limited by the use of a polynomial
of small degree, a limitation imposed by the large size of the problem.
The results presented in Table (\ref{tab:BS_3}) have similar properties
to those of Table (\ref{tab:BS_2}), however, we observe that BC loses
accuracy in this case. 

Finally, by examining Table (\ref{tab:BS_4}), we observe that, given
the same time, GPR-GHQ computation is often more efficient than LS.
This difference is particularly evident for $M=20$ and $M=30$, as
LS suffers from overfitting which is due to the small number of random
trajectories considered by the algorithm, in order to reach the fixed
computational time.

\begin{table}
\begin{centering}
\begin{tabular}{clc}
\toprule 
Symbol & Meaning & Value\tabularnewline
\midrule
$S_{0}$ & Initial spot value & $100$\tabularnewline
$r$ & Risk free i.r. & $0.05$\tabularnewline
$\sigma$ & Volatility & $0.3$\tabularnewline
$T$ & Maturity & $0.2$ or $1.0$\tabularnewline
\bottomrule
\end{tabular}
\par\end{centering}
\caption{\label{tab:BS}Parameters employed for the numerical experiments in
the Black-Scholes model.}

\end{table}

\begin{table}
\begin{centering}
\myt%
\begin{tabular}{cccccccccccccc}
\hline 
 &  & \multicolumn{7}{c}{\textbf{\footnotesize{}Longstaff-Schwartz}} &  & \multicolumn{2}{c}{{\footnotesize{}Benchmarks}} & {\footnotesize{}Berhart} & {\footnotesize{}Lelong}\tabularnewline
\cline{3-9} \cline{4-9} \cline{5-9} \cline{6-9} \cline{7-9} \cline{8-9} \cline{9-9} 
 & {\footnotesize{}$deg$} & \multicolumn{3}{c}{{\footnotesize{}$1$}} &  & \multicolumn{3}{c}{{\footnotesize{}$2$}} &  & {\footnotesize{}GPR-} & {\footnotesize{}LS} & {\footnotesize{}et al.} & \multirow{2}{*}{}\tabularnewline
\cline{3-5} \cline{4-5} \cline{5-5} \cline{7-9} \cline{8-9} \cline{9-9} 
{\footnotesize{}$M$} & {\footnotesize{}$P$} & {\footnotesize{}$10^{4}$} & {\footnotesize{}$10^{5}$} & {\footnotesize{}$10^{6}$} &  & {\footnotesize{}$10^{4}$} & {\footnotesize{}$10^{5}$} & {\footnotesize{}$10^{6}$} &  & \multicolumn{1}{c}{{\footnotesize{}GHQ}} &  &  & \tabularnewline
\hline 
{\footnotesize{}$\ensuremath{\phantom{1}}2$} &  & {\footnotesize{}$\underset{\left(1\right)}{1.87\pm0.02}$} & {\footnotesize{}$\underset{\left(1\right)}{1.88\pm0.00}$} & {\footnotesize{}$\underset{\left(5\right)}{1.87\pm0.00}$} &  & {\footnotesize{}$\underset{\left(0\right)}{1.89\pm0.02}$} & {\footnotesize{}$\underset{\left(1\right)}{1.87\pm0.00}$} & {\footnotesize{}$\underset{\left(5\right)}{1.88\pm0.00}$} &  & {\footnotesize{}$\underset{\pm0.01}{1.90}$} & {\footnotesize{}$\underset{\pm0.01}{1.89}$} & {\footnotesize{}$1.890$} & \tabularnewline
{\footnotesize{}$\ensuremath{\phantom{1}}3$} &  & {\footnotesize{}$\underset{\left(1\right)}{2.69\pm0.02}$} & {\footnotesize{}$\underset{\left(2\right)}{2.69\pm0.01}$} & {\footnotesize{}$\underset{\left(10\right)}{2.69\pm0.00}$} &  & {\footnotesize{}$\underset{\left(0\right)}{2.70\pm0.02}$} & {\footnotesize{}$\underset{\left(2\right)}{2.69\pm0.01}$} & {\footnotesize{}$\underset{\left(14\right)}{2.69\pm0.00}$} &  & {\footnotesize{}$\underset{\pm0.01}{2.68}$} & {\footnotesize{}$\underset{\pm0.01}{2.69}$} & {\footnotesize{}$2.684$} & \tabularnewline
\hline 
\end{tabular}
\par\end{centering}
\begin{centering}
\myt%
\begin{tabular}{cccccccccccccccccccc}
\hline 
 &  & \multicolumn{13}{c}{\textbf{\footnotesize{}GPR-GHQ}} &  & \multicolumn{2}{c}{{\footnotesize{}Benchmarks}} & {\footnotesize{}Berhart} & {\footnotesize{}Lelong}\tabularnewline
\cline{3-15} \cline{4-15} \cline{5-15} \cline{6-15} \cline{7-15} \cline{8-15} \cline{9-15} \cline{10-15} \cline{11-15} \cline{12-15} \cline{13-15} \cline{14-15} \cline{15-15} 
 & {\footnotesize{}$Q$} & \multicolumn{3}{c}{{\footnotesize{}$8$}} &  & \multicolumn{3}{c}{{\footnotesize{}$16$}} &  & \multicolumn{3}{c}{{\footnotesize{}$32$}} &  & {\footnotesize{}$64$} &  & \multicolumn{1}{c}{{\footnotesize{}GPR-}} & {\footnotesize{}LS} & {\footnotesize{}et al.} & \tabularnewline
\cline{3-5} \cline{4-5} \cline{5-5} \cline{7-9} \cline{8-9} \cline{9-9} \cline{11-13} \cline{12-13} \cline{13-13} \cline{15-15} 
{\footnotesize{}$M$} & {\footnotesize{}$P$} & {\footnotesize{}$250$} & {\footnotesize{}$1000$} & {\footnotesize{}$4000$} &  & {\footnotesize{}$250$} & {\footnotesize{}$1000$} & {\footnotesize{}$4000$} &  & {\footnotesize{}$250$} & {\footnotesize{}$1000$} & {\footnotesize{}$4000$} &  & {\footnotesize{}$8000$} &  & {\footnotesize{}GHQ} &  &  & \tabularnewline
\hline 
{\footnotesize{}$\ensuremath{\phantom{1}}2$} &  & \multicolumn{3}{c}{{\footnotesize{}$\underset{\left(2\right)}{1.812}$}} &  & \multicolumn{3}{c}{{\footnotesize{}$\underset{\left(2\right)}{1.873}$}} &  & \multicolumn{3}{c}{{\footnotesize{}$\underset{\left(4\right)}{1.882}$}} &  & {\footnotesize{}$\underset{\left(8\right)}{1.891}$} &  & {\footnotesize{}$\underset{\pm0.01}{1.90}$} & {\footnotesize{}$\underset{\pm0.01}{1.89}$} & {\footnotesize{}$1.890$} & \tabularnewline
\hline 
\end{tabular}
\par\end{centering}
\caption{\label{tab:BS_2}Black-Scholes model with $T=0.2$ and $N=50$ time
steps. Values in brackets are the computational times measured in
seconds.}
\end{table}

\begin{table}
\begin{centering}
\myt%
\begin{tabular}{cccccccc}
\hline 
 &  & \textbf{\footnotesize{}BC} &  & \multicolumn{2}{c}{{\footnotesize{}Benchmarks}} & {\footnotesize{}Berhart} & {\footnotesize{}Lelong}\tabularnewline
{\footnotesize{}$M$} &  &  &  & {\footnotesize{}GPR-GHQ} & {\footnotesize{}LS} & {\footnotesize{}et al.} & \tabularnewline
\hline 
{\footnotesize{}$\ensuremath{\phantom{1}}2$} &  & {\footnotesize{}$\underset{\left(1\cdot10^{-2}\right)}{0.94}$} &  & {\footnotesize{}$\underset{\pm0.01}{1.90}$} & {\footnotesize{}$\underset{\pm0.01}{1.89}$} & {\footnotesize{}$1.890$} & \tabularnewline
{\footnotesize{}$\ensuremath{\phantom{1}}3$} &  & {\footnotesize{}$\underset{\left(1\cdot10^{-2}\right)}{1.87}$} &  & {\footnotesize{}$\underset{\pm0.01}{2.68}$} & {\footnotesize{}$\underset{\pm0.01}{2.69}$} & {\footnotesize{}$2.684$} & \tabularnewline
{\footnotesize{}$\ensuremath{\phantom{1}}4$} &  & {\footnotesize{}$\underset{\left(1\cdot10^{-2}\right)}{2.75}$} &  & {\footnotesize{}$\underset{\pm0.01}{3.19}$} & {\footnotesize{}$\underset{\pm0.01}{3.19}$} & {\footnotesize{}$3.183$} & \tabularnewline
{\footnotesize{}$\ensuremath{\phantom{1}}5$} &  & {\footnotesize{}$\underset{\left(1\cdot10^{-2}\right)}{3.32}$} &  & {\footnotesize{}$\underset{\pm0.01}{3.54}$} & {\footnotesize{}$\underset{\pm0.01}{3.54}$} & {\footnotesize{}$3.526$} & {\footnotesize{}$3.531$}\tabularnewline
{\footnotesize{}$\ensuremath{\phantom{1}}6$} &  & {\footnotesize{}$\underset{\left(1\cdot10^{-2}\right)}{3.60}$} &  & {\footnotesize{}$\underset{\pm0.01}{3.80}$} & {\footnotesize{}$\underset{\pm0.01}{3.79}$} & {\footnotesize{}$3.773$} & \tabularnewline
{\footnotesize{}$\ensuremath{\phantom{1}}7$} &  & {\footnotesize{}$\underset{\left(2\cdot10^{-2}\right)}{3.86}$} &  & {\footnotesize{}$\underset{\pm0.01}{3.99}$} & {\footnotesize{}$\underset{\pm0.01}{3.99}$} & {\footnotesize{}$3.955$} & \tabularnewline
\hline 
\end{tabular}
\par\end{centering}
\caption{\label{tab:BS_2-1}Black-Scholes model with $T=0.2$ and $N=50$ time
steps. Values in brackets are the computational times measured in
seconds.}
\end{table}

\begin{table}
\begin{centering}
\mytt%
\begin{tabular}{cccccccccccc}
\hline 
 &  & \multicolumn{7}{c}{\textbf{\footnotesize{}Longstaff-Schwartz}} &  & \multicolumn{2}{c}{{\footnotesize{}Benchmarks}}\tabularnewline
\cline{3-9} \cline{4-9} \cline{5-9} \cline{6-9} \cline{7-9} \cline{8-9} \cline{9-9} 
 & {\footnotesize{}$deg$} & \multicolumn{3}{c}{{\footnotesize{}$1$}} &  & \multicolumn{3}{c}{{\footnotesize{}$2$}} &  & {\footnotesize{}GPR-} & {\footnotesize{}LS}\tabularnewline
\cline{3-5} \cline{4-5} \cline{5-5} \cline{7-9} \cline{8-9} \cline{9-9} 
{\footnotesize{}$M$} & {\footnotesize{}$P$} & {\footnotesize{}$10^{4}$} & {\footnotesize{}$10^{5}$} & {\footnotesize{}$10^{6}$} &  & {\footnotesize{}$10^{4}$} & {\footnotesize{}$10^{5}$} & {\footnotesize{}$10^{6}$} &  & {\footnotesize{}GHQ} & \tabularnewline
\hline 
{\footnotesize{}$\ensuremath{\phantom{1}}2$} &  & {\footnotesize{}$\underset{\left(1\right)}{2.29\pm0.02}$} & {\footnotesize{}$\underset{\left(3\right)}{2.30\pm0.01}$} & {\footnotesize{}$\underset{\left(26\right)}{2.30\pm0.00}$} &  & {\footnotesize{}$\underset{\left(0\right)}{2.31\pm0.02}$} & {\footnotesize{}$\underset{\left(2\right)}{2.30\pm0.01}$} & {\footnotesize{}$\underset{\left(25\right)}{2.30\pm0.00}$} &  & {\footnotesize{}$\underset{\pm0.01}{2.44}$} & {\footnotesize{}$\underset{\pm0.01}{2.44}$}\tabularnewline
{\footnotesize{}$10$} &  & {\footnotesize{}$\underset{\left(4\right)}{6.90\pm0.06}$} & {\footnotesize{}$\underset{\left(18\right)}{6.89\pm0.02}$} & {\footnotesize{}$\underset{\left(163\right)}{6.88\pm0.01}$} &  & {\footnotesize{}$\underset{\left(15\right)}{6.93\pm0.06}$} & {\footnotesize{}$\underset{\left(72\right)}{6.90\pm0.02}$} & {\footnotesize{}$\underset{\left(750\right)}{6.90\pm0.01}$} &  & {\footnotesize{}$\underset{\pm0.01}{6.90}$} & {\footnotesize{}$\underset{\pm0.01}{6.90}$}\tabularnewline
\hline 
\end{tabular}
\par\end{centering}
\begin{centering}
\mytt
\par\end{centering}
\begin{centering}
\mytt%
\begin{tabular}{cccccccccccccccccc}
\hline 
 &  & \multicolumn{13}{c}{\textbf{\footnotesize{}GPR-GHQ}} &  & \multicolumn{2}{c}{{\footnotesize{}Benchmarks}}\tabularnewline
\cline{3-15} \cline{4-15} \cline{5-15} \cline{6-15} \cline{7-15} \cline{8-15} \cline{9-15} \cline{10-15} \cline{11-15} \cline{12-15} \cline{13-15} \cline{14-15} \cline{15-15} 
 & {\footnotesize{}$G$} & \multicolumn{3}{c}{{\footnotesize{}$8$}} &  & \multicolumn{3}{c}{{\footnotesize{}$16$}} &  & \multicolumn{3}{c}{{\footnotesize{}$32$}} &  & {\footnotesize{}$64$} &  & {\footnotesize{}GPR-} & {\footnotesize{}LS}\tabularnewline
\cline{3-5} \cline{4-5} \cline{5-5} \cline{7-9} \cline{8-9} \cline{9-9} \cline{11-13} \cline{12-13} \cline{13-13} \cline{15-15} 
{\footnotesize{}$M$} & {\footnotesize{}$P$} & {\footnotesize{}$250$} & {\footnotesize{}$1000$} & {\footnotesize{}$4000$} &  & {\footnotesize{}$250$} & {\footnotesize{}$1000$} & {\footnotesize{}$4000$} &  & {\footnotesize{}$250$} & {\footnotesize{}$1000$} & {\footnotesize{}$4000$} &  & {\footnotesize{}$8000$} &  & {\footnotesize{}GHQ} & \tabularnewline
\hline 
{\footnotesize{}$\ensuremath{\phantom{1}}2$} &  & \multicolumn{3}{c}{{\footnotesize{}$\underset{\left(2\right)}{2.504}$}} &  & \multicolumn{3}{c}{{\footnotesize{}$\underset{\left(4\right)}{2.466}$}} &  & \multicolumn{3}{c}{{\footnotesize{}$\underset{\left(4\right)}{2.427}$}} &  & {\footnotesize{}$\underset{\left(18\right)}{2.436}$} &  & {\footnotesize{}$\underset{\pm0.01}{2.44}$} & {\footnotesize{}$\underset{\pm0.01}{2.44}$}\tabularnewline
\hline 
\end{tabular}
\par\end{centering}
\begin{centering}
\mytt%
\begin{tabular}{cccccc}
\hline 
 &  & \textbf{\footnotesize{}BC method} &  & \multicolumn{2}{c}{{\footnotesize{}Benchmarks}}\tabularnewline
{\footnotesize{}$M$} &  &  &  & {\footnotesize{}GPR-GHQ} & {\footnotesize{}LS}\tabularnewline
\hline 
{\footnotesize{}$\ensuremath{\phantom{1}}2$} &  & {\footnotesize{}$\underset{\left(1\cdot10^{-2}\right)}{0.94}$} &  & {\footnotesize{}$\underset{\pm0.01}{2.44}$} & {\footnotesize{}$\underset{\pm0.01}{2.44}$}\tabularnewline
{\footnotesize{}$10$} &  & {\footnotesize{}$\underset{\left(2\cdot10^{-2}\right)}{6.43}$} &  & {\footnotesize{}$\underset{\pm0.01}{6.90}$} & {\footnotesize{}$\underset{\pm0.01}{6.90}$}\tabularnewline
{\footnotesize{}$20$} &  & {\footnotesize{}$\underset{\left(2\right)}{8.49}$} &  & {\footnotesize{}$\underset{\pm0.01}{8.68}$} & {\footnotesize{}$\underset{\pm0.01}{8.67}$}\tabularnewline
{\footnotesize{}$30$} &  & {\footnotesize{}$\underset{\left(1598\right)}{9.47}$} &  & {\footnotesize{}$\underset{\pm0.01}{9.57}$} & {\footnotesize{}$\underset{\pm0.01}{9.54}$}\tabularnewline
\hline 
\end{tabular}
\par\end{centering}
\caption{\label{tab:BS_3}Black-Scholes model with $T=1.0$ and $N=250$ time
steps. Values in brackets are the computational times measured in
seconds. }
\end{table}

\begin{table}
\begin{centering}
{\footnotesize{}}%
\begin{tabular}{cccccc}
\hline 
\multicolumn{6}{c}{\textbf{\footnotesize{}Comparison }}\tabularnewline
\hline 
 & \multicolumn{2}{c}{} &  & \multicolumn{2}{c}{{\footnotesize{}Benchmarks}}\tabularnewline
{\footnotesize{}$M$} & \textbf{\footnotesize{}LS} & \textbf{\footnotesize{}GPR-GHQ} &  & {\footnotesize{}GPR-} & {\footnotesize{}LS}\tabularnewline
 & \multicolumn{2}{c}{{\footnotesize{}$30\ \text{s}$}} &  & {\footnotesize{}GHQ} & \tabularnewline
\cline{1-3} \cline{2-3} \cline{3-3} 
{\footnotesize{}$\ensuremath{\phantom{1}}2$} & {\footnotesize{}$\underset{\left(1.0e6;4\right)}{1.88\pm0.00}$} & {\footnotesize{}$\underset{\left(1;1024\right)}{1.89^{*}}$} &  & {\footnotesize{}$\underset{\pm0.01}{1.89}$} & {\footnotesize{}$\underset{\pm0.01}{1.89}$}\tabularnewline
{\footnotesize{}$10$} & {\footnotesize{}$\underset{\left(1.0e5;2\right)}{4.33^{*}\pm0.02}$} & {\footnotesize{}$\underset{\left(500;8\right)}{4.32}$} &  & {\footnotesize{}$\underset{\pm0.01}{4.34}$} & {\footnotesize{}$\underset{\pm0.01}{4.33}$}\tabularnewline
{\footnotesize{}$20$} & {\footnotesize{}$\underset{\left(2.0e4;2\right)}{4.57\pm0.06}$} & {\footnotesize{}$\underset{\left(550;4\right)}{4.47^{*}}$} &  & {\footnotesize{}$\underset{\pm0.01}{4.48}$} & {\footnotesize{}$\underset{\pm0.01}{4.47}$}\tabularnewline
{\footnotesize{}$30$} & {\footnotesize{}$\underset{\left(1.0e4;2\right)}{4.40\pm0.11}$} & {\footnotesize{}$\underset{\left(650;4\right)}{4.14^{*}}$} &  & {\footnotesize{}$\underset{\pm0.01}{4.16}$} & {\footnotesize{}$\underset{\pm0.01}{4.14}$}\tabularnewline
\hline 
\end{tabular}{\footnotesize\par}
\par\end{centering}
\caption{\label{tab:BS_4}Black-Scholes model. Comparison between the Longstaff-Schwartz
and the GPR-GHQ methods for $T=0.2$ and $N=50$. Values in brackets
are the numerical parameters: number of simulations and polynomial
degree for LS, $P$ and $Q$ for the GPR-GHQ method. Each asterisk
indicates the best value for a predetermined run-time as the closest
to the benchmarks.}
\end{table}

\subsection{The Clewlow-Strickland model}

The parameters for the Clewlow-Strickland model are the same used
by \citet{dong2021} (see Table (\ref{tab:CS})). Tables (\ref{tab:CS2})
and (\ref{tab:CS3}) report the numerical results for different values
of $M$ and for different parameter configurations. Note that both
numerical methods tend to converge to the same results. Table (\ref{tab:CS4})
reports the results for the tests with fixed computational time. The
two methods are essentially equivalent for $M\leq20$ while for $M=30$
GPR-GHQ converges much faster to the correct value as LS over-estimates
the price, suffering from overfitting.

\begin{table}
\begin{centering}
\begin{tabular}{clc}
\toprule 
Symbol & Meaning & Value\tabularnewline
\midrule
$F\left(0,t\right)$ & Initial forward curve & $100$\tabularnewline
$r$ & Risk free i.r. & $0.05$\tabularnewline
$\alpha$ & Mean reversion speed & $5$\tabularnewline
$\sigma$ & Volatility & $0.5$\tabularnewline
$T$ & Maturity & $0.2$ or $1.0$\tabularnewline
\bottomrule
\end{tabular}
\par\end{centering}
\caption{\label{tab:CS}Parameters employed for the numerical experiments in
the Clewlow-Strickland model.}
\end{table}

\begin{table}
\begin{centering}
\mytt%
\begin{tabular}{cccccccccccc}
\hline 
 &  & \multicolumn{7}{c}{\textbf{\footnotesize{}Longstaff-Schwartz}} &  & \multicolumn{2}{c}{{\footnotesize{}Benchmarks}}\tabularnewline
\cline{3-9} \cline{4-9} \cline{5-9} \cline{6-9} \cline{7-9} \cline{8-9} \cline{9-9} 
 & {\footnotesize{}$deg$} & \multicolumn{3}{c}{{\footnotesize{}$1$}} &  & \multicolumn{3}{c}{{\footnotesize{}$2$}} &  & {\footnotesize{}GPR-} & {\footnotesize{}LS}\tabularnewline
\cline{3-5} \cline{4-5} \cline{5-5} \cline{7-9} \cline{8-9} \cline{9-9} 
{\footnotesize{}$M$} & {\footnotesize{}$P$} & {\footnotesize{}$10^{4}$} & {\footnotesize{}$10^{5}$} & {\footnotesize{}$10^{6}$} &  & {\footnotesize{}$10^{4}$} & {\footnotesize{}$10^{5}$} & {\footnotesize{}$10^{6}$} &  & {\footnotesize{}GHQ} & \tabularnewline
\hline 
{\footnotesize{}$\ensuremath{\phantom{1}}2$} &  & {\footnotesize{}$\underset{\left(1\right)}{3.14\pm0.02}$} & {\footnotesize{}$\underset{\left(2\right)}{3.14\pm0.01}$} & {\footnotesize{}$\underset{\left(9\right)}{3.14\pm0.00}$} &  & {\footnotesize{}$\underset{\left(1\right)}{3.14\pm0.02}$} & {\footnotesize{}$\underset{\left(1\right)}{3.14\pm0.01}$} & {\footnotesize{}$\underset{\left(10\right)}{3.14\pm0.00}$} &  & {\footnotesize{}$\underset{\pm0.01}{3.14}$} & {\footnotesize{}$\underset{\pm0.01}{3.14}$}\tabularnewline
{\footnotesize{}$10$} &  & {\footnotesize{}$\underset{\left(1\right)}{7.26\pm0.07}$} & {\footnotesize{}$\underset{\left(4\right)}{7.24\pm0.02}$} & {\footnotesize{}$\underset{\left(26\right)}{7.23\pm0.01}$} &  & {\footnotesize{}$\underset{\left(2\right)}{7.27\pm0.07}$} & {\footnotesize{}$\underset{\left(12\right)}{7.25\pm0.02}$} & {\footnotesize{}$\underset{\left(134\right)}{7.27\pm0.01}$} &  & {\footnotesize{}$\underset{\pm0.01}{7.27}$} & {\footnotesize{}$\underset{\pm0.01}{7.27}$}\tabularnewline
{\footnotesize{}$20$} &  & {\footnotesize{}$\underset{\left(2\right)}{7.25\pm0.11}$} & {\footnotesize{}$\underset{\left(6\right)}{7.26\pm0.03}$} & {\footnotesize{}$\underset{\left(56\right)}{7.24\pm0.01}$} &  & {\footnotesize{}$\underset{\left(11\right)}{7.54\pm0.10}$} & {\footnotesize{}$\underset{\left(52\right)}{7.36\pm0.03}$} & {\footnotesize{}$\underset{\left(441\right)}{7.37\pm0.01}$} &  & {\footnotesize{}$\underset{\pm0.01}{7.37}$} & {\footnotesize{}$\underset{\pm0.01}{7.36}$}\tabularnewline
{\footnotesize{}$30$} &  & {\footnotesize{}$\underset{\left(2\right)}{6.30\pm0.12}$} & {\footnotesize{}$\underset{\left(8\right)}{6.32\pm0.04}$} & {\footnotesize{}$\underset{\left(61\right)}{6.31\pm0.01}$} &  & {\footnotesize{}$\underset{\left(24\right)}{6.83\pm0.12}$} & {\footnotesize{}$\underset{\left(104\right)}{6.45\pm0.04}$} & {\footnotesize{}$\underset{\left(1002\right)}{6.43\pm0.01}$} &  & {\footnotesize{}$\underset{\pm0.01}{6.43}$} & {\footnotesize{}$\underset{\pm0.01}{6.42}$}\tabularnewline
\hline 
\end{tabular}
\par\end{centering}
\begin{centering}
\mytt%
\begin{tabular}{cccccccccccccccccc}
\hline 
 & \multicolumn{14}{c}{\textbf{\footnotesize{}GPR-GHQ}} &  & \multicolumn{2}{c}{{\footnotesize{}Benchmarks}}\tabularnewline
\cline{3-15} \cline{4-15} \cline{5-15} \cline{6-15} \cline{7-15} \cline{8-15} \cline{9-15} \cline{10-15} \cline{11-15} \cline{12-15} \cline{13-15} \cline{14-15} \cline{15-15} 
 & {\footnotesize{}$G$} & \multicolumn{3}{c}{{\footnotesize{}$8$}} &  & \multicolumn{3}{c}{{\footnotesize{}$16$}} &  & \multicolumn{3}{c}{{\footnotesize{}$32$}} &  & {\footnotesize{}$64$} &  & {\footnotesize{}GPR-} & {\footnotesize{}LS}\tabularnewline
\cline{3-5} \cline{4-5} \cline{5-5} \cline{7-9} \cline{8-9} \cline{9-9} \cline{11-13} \cline{12-13} \cline{13-13} \cline{15-15} 
{\footnotesize{}$M$} & {\footnotesize{}$P$} & {\footnotesize{}$250$} & {\footnotesize{}$1000$} & {\footnotesize{}$4000$} &  & {\footnotesize{}$250$} & {\footnotesize{}$1000$} & {\footnotesize{}$4000$} &  & {\footnotesize{}$250$} & {\footnotesize{}$1000$} & {\footnotesize{}$4000$} &  & {\footnotesize{}$8000$} &  & {\footnotesize{}GHQ} & \tabularnewline
\hline 
{\footnotesize{}$\ensuremath{\phantom{1}}2$} &  & {\footnotesize{}$\underset{\left(32\right)}{3.003}$} & {\footnotesize{}$\underset{\left(109\right)}{3.003}$} & {\footnotesize{}$\underset{\left(443\right)}{3.003}$} &  & {\footnotesize{}$\underset{\left(34\right)}{3.109}$} & {\footnotesize{}$\underset{\left(94\right)}{3.109}$} & {\footnotesize{}$\underset{\left(391\right)}{3.109}$} &  & {\footnotesize{}$\underset{\left(32\right)}{3.124}$} & {\footnotesize{}$\underset{\left(116\right)}{3.124}$} & {\footnotesize{}$\underset{\left(430\right)}{3.124}$} &  & {\footnotesize{}$\underset{\left(1253\right)}{3.140}$} &  & {\footnotesize{}$\underset{\pm0.01}{3.14}$} & {\footnotesize{}$\underset{\pm0.01}{3.14}$}\tabularnewline
{\footnotesize{}$10$} &  & {\footnotesize{}$\underset{\left(11\right)}{7.305}$} & {\footnotesize{}$\underset{\left(59\right)}{7.281}$} & {\footnotesize{}$\underset{\left(362\right)}{7.275}$} &  & {\footnotesize{}$\underset{\left(11\right)}{7.296}$} & {\footnotesize{}$\underset{\left(65\right)}{7.278}$} & {\footnotesize{}$\underset{\left(269\right)}{7.273}$} &  & {\footnotesize{}$\underset{\left(13\right)}{7.299}$} & {\footnotesize{}$\underset{\left(75\right)}{7.274}$} & {\footnotesize{}$\underset{\left(358\right)}{7.272}$} &  & {\footnotesize{}$\underset{\left(1164\right)}{7.273}$} &  & {\footnotesize{}$\underset{\pm0.01}{7.27}$} & {\footnotesize{}$\underset{\pm0.01}{7.27}$}\tabularnewline
{\footnotesize{}$20$} &  & {\footnotesize{}$\underset{\left(9\right)}{7.414}$} & {\footnotesize{}$\underset{\left(53\right)}{7.384}$} & {\footnotesize{}$\underset{\left(203\right)}{7.378}$} &  & {\footnotesize{}$\underset{\left(10\right)}{7.411}$} & {\footnotesize{}$\underset{\left(55\right)}{7.382}$} & {\footnotesize{}$\underset{\left(220\right)}{7.377}$} &  & {\footnotesize{}$\underset{\left(11\right)}{7.409}$} & {\footnotesize{}$\underset{\left(63\right)}{7.381}$} & {\footnotesize{}$\underset{\left(491\right)}{7.377}$} &  & {\footnotesize{}$\underset{\left(956\right)}{7.376}$} &  & {\footnotesize{}$\underset{\pm0.01}{7.37}$} & {\footnotesize{}$\underset{\pm0.01}{7.36}$}\tabularnewline
{\footnotesize{}$30$} &  & {\footnotesize{}$\underset{\left(9\right)}{6.427}$} & {\footnotesize{}$\underset{\left(38\right)}{6.432}$} & {\footnotesize{}$\underset{\left(142\right)}{6.437}$} &  & {\footnotesize{}$\underset{\left(8\right)}{6.426}$} & {\footnotesize{}$\underset{\left(43\right)}{6.431}$} & {\footnotesize{}$\underset{\left(247\right)}{6.437}$} &  & {\footnotesize{}$\underset{\left(9\right)}{6.426}$} & {\footnotesize{}$\underset{\left(48\right)}{6.431}$} & {\footnotesize{}$\underset{\left(260\right)}{6.437}$} &  & {\footnotesize{}$\underset{\left(687\right)}{6.437}$} &  & {\footnotesize{}$\underset{\pm0.01}{6.43}$} & {\footnotesize{}$\underset{\pm0.01}{6.42}$}\tabularnewline
\hline 
\end{tabular}
\par\end{centering}
\caption{\label{tab:CS2}Clewlow-Strickland model with $T=0.2$ and $N=50$
time steps. Values in brackets are the computational times measured
in seconds. }
\end{table}

\begin{table}
\begin{centering}
\mytt%
\begin{tabular}{cccccccccccc}
\hline 
 &  & \multicolumn{7}{c}{\textbf{\footnotesize{}Longstaff-Schwartz}} &  & \multicolumn{2}{c}{{\footnotesize{}Benchmarks}}\tabularnewline
\cline{3-9} \cline{4-9} \cline{5-9} \cline{6-9} \cline{7-9} \cline{8-9} \cline{9-9} 
 & {\footnotesize{}$deg$} & \multicolumn{3}{c}{{\footnotesize{}$1$}} &  & \multicolumn{3}{c}{{\footnotesize{}$2$}} &  & {\footnotesize{}GPR-} & {\footnotesize{}LS}\tabularnewline
\cline{3-5} \cline{4-5} \cline{5-5} \cline{7-9} \cline{8-9} \cline{9-9} 
{\footnotesize{}$M$} & {\footnotesize{}$P$} & {\footnotesize{}$10^{4}$} & {\footnotesize{}$10^{5}$} & {\footnotesize{}$10^{6}$} &  & {\footnotesize{}$10^{4}$} & {\footnotesize{}$10^{5}$} & {\footnotesize{}$10^{6}$} &  & {\footnotesize{}GHQ} & \tabularnewline
\hline 
{\footnotesize{}$\ensuremath{\phantom{1}}2$} &  & {\footnotesize{}$\underset{\left(3\right)}{4.10\pm0.02}$} & {\footnotesize{}$\underset{\left(6\right)}{4.11\pm0.01}$} & {\footnotesize{}$\underset{\left(41\right)}{4.11\pm0.00}$} &  & {\footnotesize{}$\underset{\left(1\right)}{4.11\pm0.02}$} & {\footnotesize{}$\underset{\left(7\right)}{4.11\pm0.01}$} & {\footnotesize{}$\underset{\left(53\right)}{4.11\pm0.00}$} &  & {\footnotesize{}$\underset{\pm0.01}{4.12}$} & {\footnotesize{}$\underset{\pm0.01}{4.11}$}\tabularnewline
{\footnotesize{}$10$} &  & {\footnotesize{}$\underset{\left(4\right)}{11.73\pm0.06}$} & {\footnotesize{}$\underset{\left(17\right)}{11.75\pm0.02}$} & {\footnotesize{}$\underset{\left(160\right)}{11.73\pm0.01}$} &  & {\footnotesize{}$\underset{\left(15\right)}{11.80\pm0.06}$} & {\footnotesize{}$\underset{\left(71\right)}{11.76\pm0.02}$} & {\footnotesize{}$\underset{\left(797\right)}{11.75\pm0.01}$} &  & {\footnotesize{}$\underset{\pm0.01}{11.76}$} & {\footnotesize{}$\underset{\pm0.01}{11.75}$}\tabularnewline
{\footnotesize{}$20$} &  & {\footnotesize{}$\underset{\left(10\right)}{14.75\pm0.09}$} & {\footnotesize{}$\underset{\left(41\right)}{14.77\pm0.03}$} & {\footnotesize{}$\underset{\left(437\right)}{14.75\pm0.01}$} &  & {\footnotesize{}$\underset{\left(87\right)}{15.12\pm0.09}$} & {\footnotesize{}$\underset{\left(405\right)}{14.83\pm0.03}$} & {\footnotesize{}$\underset{\left(3531\right)}{14.80\pm0.01}$} &  & {\footnotesize{}$\underset{\pm0.01}{14.81}$} & {\footnotesize{}$\underset{\pm0.01}{14.79}$}\tabularnewline
{\footnotesize{}$30$} &  & {\footnotesize{}$\underset{\left(18\right)}{16.16\pm0.11}$} & {\footnotesize{}$\underset{\left(81\right)}{16.21\pm0.04}$} & {\footnotesize{}$\underset{\left(841\right)}{16.20\pm0.01}$} &  & {\footnotesize{}$\underset{\left(270\right)}{17.08\pm0.11}$} & {\footnotesize{}$\underset{\left(1183\right)}{16.37\pm0.03}$} & {\footnotesize{}$\underset{\left(11968\right)}{16.28\pm0.01}$} &  & {\footnotesize{}$\underset{\pm0.01}{16.30}$} & {\footnotesize{}$\underset{\pm0.01}{16.28}$}\tabularnewline
\hline 
\end{tabular}
\par\end{centering}
\begin{centering}
\mytt%
\begin{tabular}{cccccccccccccccccc}
\hline 
 &  & \multicolumn{13}{c}{\textbf{\footnotesize{}GPR-GHQ}} &  & \multicolumn{2}{c}{{\footnotesize{}Benchmarks}}\tabularnewline
\cline{3-15} \cline{4-15} \cline{5-15} \cline{6-15} \cline{7-15} \cline{8-15} \cline{9-15} \cline{10-15} \cline{11-15} \cline{12-15} \cline{13-15} \cline{14-15} \cline{15-15} 
 & {\footnotesize{}$G$} & \multicolumn{3}{c}{{\footnotesize{}$8$}} &  & \multicolumn{3}{c}{{\footnotesize{}$16$}} &  & \multicolumn{3}{c}{{\footnotesize{}$32$}} &  & {\footnotesize{}$64$} &  & {\footnotesize{}GPR-} & {\footnotesize{}LS}\tabularnewline
\cline{3-5} \cline{4-5} \cline{5-5} \cline{7-9} \cline{8-9} \cline{9-9} \cline{11-13} \cline{12-13} \cline{13-13} \cline{15-15} 
{\footnotesize{}$M$} & {\footnotesize{}$P$} & {\footnotesize{}$250$} & {\footnotesize{}$1000$} & {\footnotesize{}$4000$} &  & {\footnotesize{}$250$} & {\footnotesize{}$1000$} & {\footnotesize{}$4000$} &  & {\footnotesize{}$250$} & {\footnotesize{}$1000$} & {\footnotesize{}$4000$} &  & {\footnotesize{}$8000$} &  & {\footnotesize{}GHQ} & \tabularnewline
\hline 
{\footnotesize{}$\ensuremath{\phantom{1}}2$} &  & {\footnotesize{}$\underset{\left(50\right)}{4.214}$} & {\footnotesize{}$\underset{\left(509\right)}{4.214}$} & {\footnotesize{}$\underset{\left(1863\right)}{4.214}$} &  & {\footnotesize{}$\underset{\left(61\right)}{4.138}$} & {\footnotesize{}$\underset{\left(383\right)}{4.138}$} & {\footnotesize{}$\underset{\left(2078\right)}{4.139}$} &  & {\footnotesize{}$\underset{\left(60\right)}{4.106}$} & {\footnotesize{}$\underset{\left(433\right)}{4.107}$} & {\footnotesize{}$\underset{\left(3016\right)}{4.107}$} &  & {\footnotesize{}$\underset{\left(8759\right)}{4.110}$} &  & {\footnotesize{}$\underset{\pm0.01}{4.12}$} & {\footnotesize{}$\underset{\pm0.01}{4.11}$}\tabularnewline
{\footnotesize{}$10$} &  & {\footnotesize{}$\underset{\left(57\right)}{11.777}$} & {\footnotesize{}$\underset{\left(401\right)}{11.777}$} & {\footnotesize{}$\underset{\left(1696\right)}{11.762}$} &  & {\footnotesize{}$\underset{\left(60\right)}{11.786}$} & {\footnotesize{}$\underset{\left(537\right)}{11.772}$} & {\footnotesize{}$\underset{\left(1981\right)}{11.760}$} &  & {\footnotesize{}$\underset{\left(69\right)}{11.787}$} & {\footnotesize{}$\underset{\left(482\right)}{11.770}$} & {\footnotesize{}$\underset{\left(2475\right)}{11.759}$} &  & {\footnotesize{}$\underset{\left(7652\right)}{11.758}$} &  & {\footnotesize{}$\underset{\pm0.01}{11.76}$} & {\footnotesize{}$\underset{\pm0.01}{11.75}$}\tabularnewline
{\footnotesize{}$20$} &  & {\footnotesize{}$\underset{\left(86\right)}{15.124}$} & {\footnotesize{}$\underset{\left(508\right)}{14.887}$} & {\footnotesize{}$\underset{\left(2585\right)}{14.835}$} &  & {\footnotesize{}$\underset{\left(92\right)}{15.125}$} & {\footnotesize{}$\underset{\left(637\right)}{14.888}$} & {\footnotesize{}$\underset{\left(2065\right)}{14.836}$} &  & {\footnotesize{}$\underset{\left(104\right)}{15.124}$} & {\footnotesize{}$\underset{\left(542\right)}{14.887}$} & {\footnotesize{}$\underset{\left(3035\right)}{14.836}$} &  & {\footnotesize{}$\underset{\left(7486\right)}{14.826}$} &  & {\footnotesize{}$\underset{\pm0.01}{14.81}$} & {\footnotesize{}$\underset{\pm0.01}{14.79}$}\tabularnewline
{\footnotesize{}$30$} &  & {\footnotesize{}$\underset{\left(74\right)}{16.718}$} & {\footnotesize{}$\underset{\left(551\right)}{16.408}$} & {\footnotesize{}$\underset{\left(1936\right)}{16.333}$} &  & {\footnotesize{}$\underset{\left(76\right)}{16.720}$} & {\footnotesize{}$\underset{\left(494\right)}{16.409}$} & {\footnotesize{}$\underset{\left(2144\right)}{16.332}$} &  & {\footnotesize{}$\underset{\left(93\right)}{16.721}$} & {\footnotesize{}$\underset{\left(558\right)}{16.409}$} & {\footnotesize{}$\underset{\left(2235\right)}{16.332}$} &  & {\footnotesize{}$\underset{\left(7067\right)}{16.325}$} &  & {\footnotesize{}$\underset{\pm0.01}{16.30}$} & {\footnotesize{}$\underset{\pm0.01}{16.28}$}\tabularnewline
\hline 
\end{tabular}
\par\end{centering}
\caption{\label{tab:CS3}Clewlow-Strickland model with $T=1.0$ and $N=250$
time steps. Values in brackets are the computational times measured
in seconds. }
\end{table}

\begin{table}
\begin{centering}
\myttt
\par\end{centering}
\begin{centering}
\begin{tabular}{cccccc}
\hline 
\multicolumn{6}{c}{\textbf{\footnotesize{}Comparison}}\tabularnewline
\hline 
 & \multicolumn{2}{c}{} &  & \multicolumn{2}{c}{{\footnotesize{}Benchmarks}}\tabularnewline
{\footnotesize{}$M$} & {\footnotesize{}LSB} & {\footnotesize{}GPR} &  & {\footnotesize{}GPR-} & {\footnotesize{}LS}\tabularnewline
\cline{1-3} \cline{2-3} \cline{3-3} 
 & \multicolumn{2}{c}{{\footnotesize{}$30\ \text{s}$}} &  & {\footnotesize{}GHQ} & \tabularnewline
\cline{2-3} \cline{3-3} 
{\footnotesize{}$\ensuremath{\phantom{1}}2$} & {\footnotesize{}$\underset{\left(1.0e6;2\right)}{3.14^{*}\pm0.00}$} & {\footnotesize{}$\underset{\left(450;64\right)}{3.14^{*}}$} &  & {\footnotesize{}$\underset{\pm0.01}{3.14}$} & {\footnotesize{}$\underset{\pm0.01}{3.14}$}\tabularnewline
{\footnotesize{}$10$} & {\footnotesize{}$\underset{\left(1.0e5;2\right)}{7.28^{*}\pm0.02}$} & {\footnotesize{}$\underset{\left(550;16\right)}{7.28^{*}}$} &  & {\footnotesize{}$\underset{\pm0.01}{7.27}$} & {\footnotesize{}$\underset{\pm0.01}{7.27}$}\tabularnewline
{\footnotesize{}$20$} & {\footnotesize{}$\underset{\left(2.5e4;2\right)}{7.41\pm0.06}$} & {\footnotesize{}$\underset{\left(600;4\right)}{7.38^{*}}$} &  & {\footnotesize{}$\underset{\pm0.01}{7.37}$} & {\footnotesize{}$\underset{\pm0.01}{7.36}$}\tabularnewline
{\footnotesize{}$30$} & {\footnotesize{}$\underset{\left(1.0e4;2\right)}{6.88\pm0.12}$} & {\footnotesize{}$\underset{\left(750;4\right)}{6.44^{*}}$} &  & {\footnotesize{}$\underset{\pm0.01}{6.43}$} & {\footnotesize{}$\underset{\pm0.01}{6.42}$}\tabularnewline
\hline 
 &  &  &  &  & \tabularnewline
 & \multicolumn{2}{c}{{\footnotesize{}$1\ \text{min}$}} &  &  & \tabularnewline
\cline{2-3} \cline{3-3} 
{\footnotesize{}$\ensuremath{\phantom{1}}2$} & {\footnotesize{}$\underset{\left(1.0e6;2\right)}{3.14^{*}\pm0.00}$} & {\footnotesize{}$\underset{\left(600;64\right)}{3.14^{*}}$} &  & {\footnotesize{}$\underset{\pm0.01}{3.14}$} & {\footnotesize{}$\underset{\pm0.01}{3.14}$}\tabularnewline
{\footnotesize{}$10$} & {\footnotesize{}$\underset{\left(1.7e5;2\right)}{7.28^{*}\pm0.02}$} & {\footnotesize{}$\underset{\left(800;16\right)}{7.28^{*}}$} &  & {\footnotesize{}$\underset{\pm0.01}{7.27}$} & {\footnotesize{}$\underset{\pm0.01}{7.27}$}\tabularnewline
{\footnotesize{}$20$} & {\footnotesize{}$\underset{\left(4.8e4;2\right)}{7.43\pm0.04}$} & {\footnotesize{}$\underset{\left(900;6\right)}{7.38^{*}}$} &  & {\footnotesize{}$\underset{\pm0.01}{7.37}$} & {\footnotesize{}$\underset{\pm0.01}{7.36}$}\tabularnewline
{\footnotesize{}$30$} & {\footnotesize{}$\underset{\left(2.0e4;2\right)}{6.77\pm0.09}$} & {\footnotesize{}$\underset{\left(1000;8\right)}{6.43^{*}}$} &  & {\footnotesize{}$\underset{\pm0.01}{6.43}$} & {\footnotesize{}$\underset{\pm0.01}{6.42}$}\tabularnewline
\hline 
 &  &  &  &  & \tabularnewline
 & \multicolumn{2}{c}{{\footnotesize{}$2\ \text{min}$}} &  &  & \tabularnewline
\cline{2-3} \cline{3-3} 
{\footnotesize{}$\ensuremath{\phantom{1}}2$} & {\footnotesize{}$\underset{\left(1.0e6;2\right)}{3.14^{*}\pm0.00}$} & {\footnotesize{}$\underset{\left(900;64\right)}{3.14^{*}}$} &  & {\footnotesize{}$\underset{\pm0.01}{3.14}$} & {\footnotesize{}$\underset{\pm0.01}{3.14}$}\tabularnewline
{\footnotesize{}$10$} & {\footnotesize{}$\underset{\left(3.0e5;2\right)}{7.27^{*}\pm0.01}$} & {\footnotesize{}$\underset{\left(900;32\right)}{7.28}$} &  & {\footnotesize{}$\underset{\pm0.01}{7.27}$} & {\footnotesize{}$\underset{\pm0.01}{7.27}$}\tabularnewline
{\footnotesize{}$20$} & {\footnotesize{}$\underset{\left(9.0e4;2\right)}{7.37^{*}\pm0.03}$} & {\footnotesize{}$\underset{\left(1100;6\right)}{7.38}$} &  & {\footnotesize{}$\underset{\pm0.01}{7.37}$} & {\footnotesize{}$\underset{\pm0.01}{7.36}$}\tabularnewline
{\footnotesize{}$30$} & {\footnotesize{}$\underset{\left(3.9e4;2\right)}{6.55\pm0.06}$} & {\footnotesize{}$\underset{\left(1300;8\right)}{6.43^{*}}$} &  & {\footnotesize{}$\underset{\pm0.01}{6.43}$} & {\footnotesize{}$\underset{\pm0.01}{6.42}$}\tabularnewline
\hline 
\end{tabular}
\par\end{centering}
\caption{\label{tab:CS4}Clewlow-Strickland model. Comparison between the Longstaff-Schwartz
and the GPR-GHQ methods for $T=0.2$ and $N=50$. Values in brackets
are the numerical parameters: number of simulations and polynomial
degree for LS, $P$ and $Q$ for the GPR-GHQ method. Each asterisk
indicates the best value for a predetermined run-time as the closest
to the benchmarks.}
\end{table}

\subsection{The rough-Bergomi model}

The parameters for the rough-Bergomi model are the same used by \citet{bayer2020}
(see Table (\ref{tab:rB})). Tables (\ref{tab:rB2}) and (\ref{tab:rB3})
report the numerical results for different values of $M$ and for
different parameter configurations. We observe that, in case, the
values provided by LS are generally a few cents lower than the benchmarks,
which indicates that convergence is slower in the rough-Bergomi model
than in the Black-Scholes and Clewlow-Strickland models. Conversely,
the values returned by GPR-GHQ are very close to the benchmark values
even using small values for $P$ and $Q$. We then observe that, both
for LS and for GPR-GHQ, they tend to grow with increasing $J$. This
is consistent with the fact that having more information, it is possible
to better learn the continuation value and therefore improve the exercise
strategy. Obviously, the increase in computational times is the negative
aspect that weighs on the choice of large values for $J$. Please
note that the differences between $J=7$ and $J=15$ are of the order
of at most 1 cent, therefore we used $J=15$ for the calculation of
the benchmark.

Table (\ref{tab:rB4}) reports the results for the tests with fixed
computational time. GPR-GHQ is very often the best method, in particular
for $M=20$ and $M=3$0, as well as when the computational target
time is 2 minutes. It should be noted that the calculation of the
price of a moving average option in the rough-Bergomi model has a
greater dimension than the same problem in the Black-Scholes and Clewlow-Strickland
models since volatility history must also be included in the set for
predictors. Consequently, GPR-GHQ, is particularly efficient to tackle
this type of problem.

\begin{table}
\begin{centering}
\begin{tabular}{clc}
\toprule 
Symbol & Meaning & Value\tabularnewline
\midrule
$S_{0}$ & Initial spot value & $100$\tabularnewline
$r$ & Risk free i.r. & $0.05$\tabularnewline
$\rho$ & correlation & $-0.9$\tabularnewline
$\xi_{0}$ & Forward variance rate & $0.09$\tabularnewline
$\eta$ & Vol of vol & $1.9$\tabularnewline
$T$ & Maturity & $0.2$ or $1.0$\tabularnewline
\bottomrule
\end{tabular}
\par\end{centering}
\caption{\label{tab:rB}Parameters employed for the numerical experiments in
the Clewlow-Strickland model.}
\end{table}

\begin{table}
\begin{centering}
\mytt%
\begin{tabular}{cccccccccccc}
\hline 
 &  & \multicolumn{7}{c}{\textbf{\footnotesize{}Longstaff-Schwartz}} &  &  & \tabularnewline
\cline{3-9} \cline{4-9} \cline{5-9} \cline{6-9} \cline{7-9} \cline{8-9} \cline{9-9} 
 & {\footnotesize{}$deg$} & \multicolumn{3}{c}{{\footnotesize{}$1$}} &  & \multicolumn{3}{c}{{\footnotesize{}$2$}} &  & \multicolumn{2}{c}{{\footnotesize{}Benchmarks}}\tabularnewline
\cline{3-5} \cline{4-5} \cline{5-5} \cline{7-9} \cline{8-9} \cline{9-9} 
 & {\footnotesize{}$P$} & \multicolumn{3}{c}{{\footnotesize{}$10^{5}$}} &  & \multicolumn{3}{c}{{\footnotesize{}$10^{6}$}} &  & {\footnotesize{}GPR-} & {\footnotesize{}LS}\tabularnewline
\cline{3-5} \cline{4-5} \cline{5-5} \cline{7-9} \cline{8-9} \cline{9-9} 
{\footnotesize{}$M$} & {\footnotesize{}$J$} & {\footnotesize{}$3$} & {\footnotesize{}$7$} & {\footnotesize{}$15$} &  & {\footnotesize{}$3$} & {\footnotesize{}$7$} & {\footnotesize{}$15$} &  & {\footnotesize{}GHQ} & \tabularnewline
\hline 
{\footnotesize{}$\ensuremath{\phantom{1}}2$} &  & {\footnotesize{}$\underset{\left(4\right)}{2.13\pm0.01}$} & {\footnotesize{}$\underset{\left(6\right)}{2.14\pm0.01}$} & {\footnotesize{}$\underset{\left(17\right)}{2.14\pm0.01}$} &  & {\footnotesize{}$\underset{\left(94\right)}{2.14\pm0.00}$} & {\footnotesize{}$\underset{\left(477\right)}{2.14\pm0.00}$} & {\footnotesize{}$\underset{\left(2500\right)}{2.15\pm0.00}$} &  & {\footnotesize{}$\underset{\pm0.01}{2.17}$} & {\footnotesize{}$\underset{\pm0.01}{2.16}$}\tabularnewline
{\footnotesize{}$10$} &  & {\footnotesize{}$\underset{\left(5\right)}{3.42\pm0.02}$} & {\footnotesize{}$\underset{\left(8\right)}{3.44\pm0.02}$} & {\footnotesize{}$\underset{\left(16\right)}{3.44\pm0.02}$} &  & {\footnotesize{}$\underset{\left(346\right)}{3.49\pm0.00}$} & {\footnotesize{}$\underset{\left(561\right)}{3.50\pm0.00}$} & {\footnotesize{}$\underset{\left(3154\right)}{3.50\pm0.00}$} &  & {\footnotesize{}$\underset{\pm0.01}{3.57}$} & {\footnotesize{}$\underset{\pm0.01}{3.56}$}\tabularnewline
\hline 
\end{tabular}
\par\end{centering}
\begin{centering}
\mytt%
\begin{tabular}{cccccccccccccccccccc}
\hline 
 &  & \multicolumn{15}{c}{\textbf{\footnotesize{}GPR-GHQ}} &  &  & \tabularnewline
\cline{3-17} \cline{4-17} \cline{5-17} \cline{6-17} \cline{7-17} \cline{8-17} \cline{9-17} \cline{10-17} \cline{11-17} \cline{12-17} \cline{13-17} \cline{14-17} \cline{15-17} \cline{16-17} \cline{17-17} 
 & {\footnotesize{}$G$} & \multicolumn{7}{c}{{\footnotesize{}$8$}} &  & \multicolumn{7}{c}{{\footnotesize{}$16$}} &  & \multicolumn{2}{c}{{\footnotesize{}Benchmarks}}\tabularnewline
\cline{3-9} \cline{4-9} \cline{5-9} \cline{6-9} \cline{7-9} \cline{8-9} \cline{9-9} \cline{11-17} \cline{12-17} \cline{13-17} \cline{14-17} \cline{15-17} \cline{16-17} \cline{17-17} 
 & {\footnotesize{}$P$} & \multicolumn{3}{c}{{\footnotesize{}$4000$}} &  & \multicolumn{3}{c}{{\footnotesize{}$8000$}} &  & \multicolumn{3}{c}{{\footnotesize{}$4000$}} &  & \multicolumn{3}{c}{{\footnotesize{}$8000$}} &  & {\footnotesize{}GPR-} & {\footnotesize{}LS}\tabularnewline
\cline{3-5} \cline{4-5} \cline{5-5} \cline{7-9} \cline{8-9} \cline{9-9} \cline{11-13} \cline{12-13} \cline{13-13} \cline{15-17} \cline{16-17} \cline{17-17} 
{\footnotesize{}$M$} & {\footnotesize{}$J$} & {\footnotesize{}$3$} & {\footnotesize{}$7$} & {\footnotesize{}$15$} &  & {\footnotesize{}$3$} & {\footnotesize{}$7$} & {\footnotesize{}$15$} &  & {\footnotesize{}$3$} & {\footnotesize{}$7$} & {\footnotesize{}$15$} &  & {\footnotesize{}$3$} & {\footnotesize{}$7$} & {\footnotesize{}$15$} &  & {\footnotesize{}GHQ} & \tabularnewline
\hline 
\end{tabular}
\par\end{centering}
\caption{\label{tab:rB2}rough-Bergomi model with $T=0.2$ and $N=50$ time
steps. Values in brackets are the computational times measured in
seconds. }
\end{table}

\begin{table}
\begin{centering}
\mytt%
\begin{tabular}{cccccccccccc}
\hline 
 &  & \multicolumn{7}{c}{\textbf{\footnotesize{}Longstaff-Schwartz}} &  &  & \tabularnewline
\cline{3-9} \cline{4-9} \cline{5-9} \cline{6-9} \cline{7-9} \cline{8-9} \cline{9-9} 
 & {\footnotesize{}$deg$} & \multicolumn{3}{c}{{\footnotesize{}$1$}} &  & \multicolumn{3}{c}{{\footnotesize{}$2$}} &  & \multicolumn{2}{c}{{\footnotesize{}Benchmarks}}\tabularnewline
\cline{3-5} \cline{4-5} \cline{5-5} \cline{7-9} \cline{8-9} \cline{9-9} 
 & {\footnotesize{}$P$} & \multicolumn{3}{c}{{\footnotesize{}$10^{5}$}} &  & \multicolumn{3}{c}{{\footnotesize{}$10^{6}$}} &  & {\footnotesize{}GPR-} & {\footnotesize{}LS}\tabularnewline
\cline{3-5} \cline{4-5} \cline{5-5} \cline{7-9} \cline{8-9} \cline{9-9} 
{\footnotesize{}$M$} & {\footnotesize{}$J$} & {\footnotesize{}$3$} & {\footnotesize{}$7$} & {\footnotesize{}$15$} &  & {\footnotesize{}$3$} & {\footnotesize{}$7$} & {\footnotesize{}$15$} &  & {\footnotesize{}GHQ} & \tabularnewline
\hline 
{\footnotesize{}$\ensuremath{\phantom{1}}2$} &  & {\footnotesize{}$\underset{\left(14\right)}{3.16\pm0.01}$} & {\footnotesize{}$\underset{\left(42\right)}{3.16\pm0.01}$} & {\footnotesize{}$\underset{\left(93\right)}{3.17\pm0.01}$} &  & {\footnotesize{}$\underset{\left(511\right)}{3.16\pm0.00}$} & {\footnotesize{}$\underset{\left(2135\right)}{3.19\pm0.00}$} & {\footnotesize{}$\underset{\left(14111\right)}{3.21\pm0.00}$} &  & {\footnotesize{}$\underset{\pm0.01}{3.27}$} & {\footnotesize{}$\underset{\pm0.01}{3.28}$}\tabularnewline
{\footnotesize{}$10$} &  & {\footnotesize{}$\underset{\left(28\right)}{5.07\pm0.02}$} & {\footnotesize{}$\underset{\left(54\right)}{5.09\pm0.02}$} & {\footnotesize{}$\underset{\left(100\right)}{5.10\pm0.02}$} &  & {\footnotesize{}$\underset{\left(1732\right)}{5.21\pm0.01}$} & {\footnotesize{}$\underset{\left(2983\right)}{5.24\pm0.01}$} & {\footnotesize{}$\underset{\left(15337\right)}{5.28\pm0.01}$} &  & {\footnotesize{}$\underset{\pm0.01}{5.42}$} & {\footnotesize{}$\underset{\pm0.01}{5.42}$}\tabularnewline
\hline 
\end{tabular}
\par\end{centering}
\begin{centering}
\mytt%
\begin{tabular}{cccccccccccccccccccc}
\hline 
 &  & \multicolumn{15}{c}{\textbf{\footnotesize{}GPR-GHQ}} &  &  & \tabularnewline
\cline{3-17} \cline{4-17} \cline{5-17} \cline{6-17} \cline{7-17} \cline{8-17} \cline{9-17} \cline{10-17} \cline{11-17} \cline{12-17} \cline{13-17} \cline{14-17} \cline{15-17} \cline{16-17} \cline{17-17} 
 & {\footnotesize{}$G$} & \multicolumn{7}{c}{{\footnotesize{}$8$}} &  & \multicolumn{7}{c}{{\footnotesize{}$16$}} &  & \multicolumn{2}{c}{{\footnotesize{}Benchmarks}}\tabularnewline
\cline{3-9} \cline{4-9} \cline{5-9} \cline{6-9} \cline{7-9} \cline{8-9} \cline{9-9} \cline{11-17} \cline{12-17} \cline{13-17} \cline{14-17} \cline{15-17} \cline{16-17} \cline{17-17} 
 & {\footnotesize{}$P$} & \multicolumn{3}{c}{{\footnotesize{}$4000$}} &  & \multicolumn{3}{c}{{\footnotesize{}$8000$}} &  & \multicolumn{3}{c}{{\footnotesize{}$4000$}} &  & \multicolumn{3}{c}{{\footnotesize{}$8000$}} &  & {\footnotesize{}GPR-} & {\footnotesize{}LS}\tabularnewline
\cline{3-5} \cline{4-5} \cline{5-5} \cline{7-9} \cline{8-9} \cline{9-9} \cline{11-13} \cline{12-13} \cline{13-13} \cline{15-17} \cline{16-17} \cline{17-17} 
{\footnotesize{}$M$} & {\footnotesize{}$J$} & {\footnotesize{}$3$} & {\footnotesize{}$7$} & {\footnotesize{}$15$} &  & {\footnotesize{}$3$} & {\footnotesize{}$7$} & {\footnotesize{}$15$} &  & {\footnotesize{}$3$} & {\footnotesize{}$7$} & {\footnotesize{}$15$} &  & {\footnotesize{}$3$} & {\footnotesize{}$7$} & {\footnotesize{}$15$} &  & {\footnotesize{}GHQ} & \tabularnewline
\hline 
\end{tabular}
\par\end{centering}
\caption{\label{tab:rB3}rough-Bergomi model with $T=1.0$ and $N=250$ time
steps. Values in brackets are the computational times measured in
seconds. }
\end{table}

\begin{table}
\begin{centering}
\myttt
\par\end{centering}
\begin{centering}
{\footnotesize{}}%
\begin{tabular}{cccccc}
\hline 
\multicolumn{6}{c}{\textbf{\footnotesize{}Comparison}}\tabularnewline
\hline 
 & \multicolumn{2}{c}{} &  & \multicolumn{2}{c}{{\footnotesize{}Benchmarks}}\tabularnewline
{\footnotesize{}$M$} & {\footnotesize{}LSB} & {\footnotesize{}GPR} &  & {\footnotesize{}GPR-} & {\footnotesize{}LS}\tabularnewline
 & \multicolumn{2}{c}{{\footnotesize{}$30\ \text{s}$}} &  & {\footnotesize{}GHQ} & \tabularnewline
\cline{2-3} \cline{3-3} 
{\footnotesize{}$\ensuremath{\phantom{1}}2$} & {\footnotesize{}$\underset{\left(2.6e4;2\right)}{2.16^{*}\pm0.02}$} & {\footnotesize{}$\underset{\left(400;6\right)}{2.14}$} &  & {\footnotesize{}$\underset{\pm0.01}{2.17}$} & {\footnotesize{}$\underset{\pm0.01}{2.16}$}\tabularnewline
{\footnotesize{}$10$} & {\footnotesize{}$\underset{\left(2.0e4;2\right)}{3.52^{*}\pm0.03}$} & {\footnotesize{}$\underset{\left(550;4\right)}{3.52^{*}}$} &  & {\footnotesize{}$\underset{\pm0.01}{3.57}$} & {\footnotesize{}$\underset{\pm0.01}{3.56}$}\tabularnewline
{\footnotesize{}$20$} & {\footnotesize{}$\underset{\left(7.0e3;2\right)}{3.84\pm0.07}$} & {\footnotesize{}$\underset{\left(600;4\right)}{3.67^{*}}$} &  & {\footnotesize{}$\underset{\pm0.01}{3.67}$} & {\footnotesize{}$\underset{\pm0.01}{3.65}$}\tabularnewline
{\footnotesize{}$30$} & {\footnotesize{}$\underset{\left(4.4e3;2\right)}{3.85\pm0.09}$} & {\footnotesize{}$\underset{\left(600;4\right)}{3.49^{*}}$} &  & {\footnotesize{}$\underset{\pm0.01}{3.44}$} & {\footnotesize{}$\underset{\pm0.01}{3.42}$}\tabularnewline
\hline 
\end{tabular}{\footnotesize\par}
\par\end{centering}
\caption{\label{tab:rB4}rough-Bergomi model. Comparison between the Longstaff-Schwartz
and the GPR-GHQ methods for $T=0.2$, $N=50$ and $J=7.$ Values in
brackets are the numerical parameters: number of simulations and polynomial
degree for LS, $P$ and $Q$ for the GPR-GHQ method. Each asterisk
indicates the best value for a predetermined run-time as the closest
to the benchmarks.}
\end{table}

\section{Conclusion}

In this paper, we have discussed the problem of calculating the price
of a moving average option. The problem is particularly interesting
as this type of options is used in corporate finance and in the energy
commodities market. Traditional Longstaff-Schwartz methods are not
particularly efficient when the averaging window includes a few dozen
observations because the size of the problem is high. For the same
reason, tree methods are also not a general solution to this type
of problem. To solve this problem, we have proposed an innovative
method that exploits both the Machine Learning technique known as
GPR and the classical Gauss-Hermite quadrature technique. The method
is made even more efficient by a series of observations, such as the
use of some particular predictors for learning the continuation value
and by the similarity reduction in the case of the Black-Scholes model.
Similarity reduction has also been exploited to propose an improved
version of the binomial tree algorithm already present in the literature.
The proposed method has been tested in three stochastic models and
has proven successful over Longstaff-Schwartz especially when a long
window is considered. The method was also particularly effective in
the rough-Bergomi model, which considers stochastic volatility with
memory. Numerical tests have demonstrated the goodness of the proposed
approach and its convenience compared to the Longstaff-Schwartz method.
In conclusion, the proposed method is reliable and efficient for the
evaluation of moving average options.

\FloatBarrier

\bibliographystyle{apalike}
\bibliography{bibliography}

\appendix

\section{Simulation of the rough-Bergomi model \label{sec:Simulation-of-Stochastic}}

It is not possible to exactly simulate the rough-Bergomi model, however
a good approximation can be obtained by employing the discrete simulation
scheme introduced by Bayer et al. \citet{bayer2016}, that is simulating
the couple $\left(S_{t},V_{t}\right)$ on a finite number $N$ of
dates $\left\{ t_{n}=n\,\Delta t\right\} _{n=0,\dots,N}$ with $\Delta t=\frac{T}{N}$
the time increment. Specifically, if we set $\Delta W_{n}^{1}=W_{t_{n}}^{1}-W_{t_{n-1}}^{1}$,
then the $2N$-dimensional random vector $\mathbf{R}$, given by 
\begin{equation}
\mathbf{R}=\left(\Delta W_{1}^{1},\widetilde{W}_{t_{1}}^{H},\dots,\Delta W_{N}^{1},\widetilde{W}_{t_{N}}^{H}\right)^{\top},
\end{equation}
 follows a zero-mean Gaussian distribution with covariances reported
in Table \ref{tab:Formulas-for-computing}.
\begin{table}
\begin{centering}
\setlength{\tabcolsep}{4pt} \renewcommand{\arraystretch}{1.25}%
\begin{tabular}{ccccc}
\toprule 
$\mathrm{Cov}\left(\cdot,\cdot\right)$ &  & $\Delta W_{n}^{1}$ & $\widetilde{W}_{t_{n}}^{H}$ & $\widetilde{W}_{t_{m}}^{H}$\tabularnewline
\cmidrule{2-5} \cmidrule{3-5} \cmidrule{4-5} \cmidrule{5-5} 
$\Delta W_{n}^{1}$ &  & {\small{}$\Delta t$} & {\small{}$\frac{2\rho\sqrt{2H}}{2H+1}\left(\Delta t\right)^{H+\frac{1}{2}}$} & {\small{}$0$}\tabularnewline
\cmidrule{2-5} \cmidrule{3-5} \cmidrule{4-5} \cmidrule{5-5} 
$\widetilde{W}_{t_{n}}^{H}$ &  & {\small{}$\frac{2\rho\sqrt{2H}}{2H+1}\left(\Delta t\right)^{H+\frac{1}{2}}$} & {\small{}$\left(t_{n}\right)^{2H}$} & {\small{}$2H\left(t_{m}\right)^{2H}\cdot\int_{0}^{1}\frac{ds}{\left(1-s\right)^{\frac{1}{2}-H}\left(\frac{t_{n}}{t_{m}}-s\right)^{\frac{1}{2}-H}}$}\tabularnewline
\cmidrule{2-5} \cmidrule{3-5} \cmidrule{4-5} \cmidrule{5-5} 
$\Delta W_{m}^{1}$ &  & {\small{}$0$} & {\small{}$\frac{2\rho\sqrt{2H}}{2H+1}\left(\left(t_{n}-t_{m-1}\right)^{H+\frac{1}{2}}-\left(t_{n}-t_{m}\right)^{H+\frac{1}{2}}\right)$} & {\small{}$\frac{2\rho\sqrt{2H}}{2H+1}\left(\Delta t\right)^{H+\frac{1}{2}}$}\tabularnewline
\bottomrule
\end{tabular}
\par\end{centering}
\caption{\label{tab:Formulas-for-computing}Covariances about the components
of the Gaussian vector $\mathbf{R}$ for $t_{m}<t_{n}$.}

\end{table}
By using the Cholesky factorization one can compute $\Lambda$, the
lower triangular square root of the covariance matrix of $\mathbf{R}$.
Let $\mathbf{G}=\left(G_{1},\dots,G_{2N}\right)^{\top}$ be a random
vector of independent standard Gaussian random variables. Then, $\Lambda\mathbf{G}$
has the same law of the vector $\mathbf{R}$. Finally, a simulation
for $\left(S_{t_{n}},V_{t_{n}}\right)_{n=0,\dots,N}$ can be obtained
from $\mathbf{R}$ through the Euler-Maruyama scheme given by 
\begin{equation}
\begin{cases}
S_{t_{n+1}} & =S_{t_{n}}\exp\left(\left(r-\frac{1}{2}V_{t_{n}}\right)\Delta t+\sqrt{V_{t_{n}}}\Delta W_{n+1}^{1}\right)\\
V_{t_{n+1}} & =\xi_{0}\exp\left(-\frac{1}{2}\eta^{2}\left(t_{n+1}\right)^{2H}+\eta\widetilde{W}_{t_{n+1}}^{H}\right),
\end{cases}\label{eq:EM}
\end{equation}
and the initial values 
\begin{equation}
S_{t_{0}}=S_{0},\ V_{t_{0}}=\xi_{0}.\label{eq:62}
\end{equation}

\section{Gaussian Process Regression\label{sec:Gaussian-Process-Regression}}

GPR is a supervised Machine Learning technique used for regressions.
Here we report some details and we refer the interested reader to
the seminal book of \citet{rasmussen2006}.

Let us suppose that a set of $P$ couples $\mathcal{D}=\left\{ \left(\mathbf{x}_{i},y_{i}\right),i=1,\dots,P\right\} \subset\mathbb{R}^{D}\times\mathbb{R}$
is given. The $y$ observations are modeled as the realization of
the sum of a Gaussian process $\mathcal{G}$ and a Gaussian noise
source $\varepsilon$ at the $\mathbf{x}$ points. In particular,
the distribution of the vector $\mathbf{y}=\left(y_{1}\dots y_{P}\right)$
is assumed to be given by 
\begin{equation}
\mathbf{y}\sim\mathcal{N}\left(\mu\left(X\right),K\left(X,X\right)+\sigma_{P}^{2}I_{P}\right),
\end{equation}
with $\mu$ the mean function, $I_{P}$ the $P\times P$ identity
matrix and $K$ a $P\times P$ matrix given by $K\left(X,X\right)_{i,j}=k\left(\mathbf{x}_{i},\mathbf{x}_{j}\right)$
with $k:\mathbb{R}^{D}\times\mathbb{R}^{D}\rightarrow\mathbb{R}$
a certain function termed Kernel function. Many choices are possible
for $k$, but we consider the Squared Exponential kernel, which is
a $\mathcal{C}^{\infty}$ function given by
\begin{equation}
k\left(\mathbf{x},\mathbf{x}'\right)=\sigma_{f}^{2}\exp\left(-\frac{1}{2}\sum_{k=1}^{D}\frac{1}{\sigma_{l}^{2}}\left(x_{k}-x_{k}'\right)^{2}\right),
\end{equation}
where $\sigma_{f}^{2}$ is the signal variance and $\sigma_{l}^{2}$
is the length-scale.

The main purpose of GPR is extrapolate form $\mathcal{D}$, so let
us consider a test set $\tilde{X}$ of $m$ points $\left\{ \tilde{\mathbf{x}}_{j}|j=1,\dots,m\right\} $.
The realizations $\tilde{f}_{j}=\mathcal{G}\left(\tilde{\mathbf{x}}_{j}\right)+\varepsilon_{j}$
are not known but are predicted through $\mathbb{E}\left[\mathbf{\tilde{f}}|\tilde{X},\mathbf{y},X\right]$:
\begin{align}
\mathbb{E}\left[\mathbf{\tilde{f}}|\tilde{X},\mathbf{y},X\right] & =\mu\left(\tilde{X}\right)+K\left(\tilde{X},X\right)\varTheta,\label{eq:GPR_prediction}
\end{align}
with $\varTheta=\left[K\left(X,X\right)+\sigma_{P}^{2}I_{P}\right]^{-1}\left(\mathbf{y}-\mu\left(X\right)\right)$,
a $P\times1$ vector which does not depend on $\tilde{X}$. For our
purposes, we consider the mean function $\mu$ as a linear function
of the predictors and we estimate it by means of least squares regression.
Finally, log likelihood maximization is employed to estimate the so
called hyperparameters namely $\sigma_{f}^{2}$, $\sigma_{l}^{2}$
of the kernel and $\sigma_{n}^{2}$ of the noise.

The GPR method is employed in two steps: training and evaluation (also
called testing). During the training step, the function $\mu$ is
estimated together with the hyperparameters and the vector $\varTheta$
is computed. During the evaluation step, the predictions are obtained
by means of equation (\ref{eq:GPR_prediction}).

Although GPR is recognized as an accurate and efficient method when
the observed sample size $P$ is small, the method is not particularly
recommended for large samples. In fact, the overall calculation complexity
is $O\left(P^{3}\right)$ and the memory consumption is $O\left(P^{2}\right)$.
Due to these limitations, the value for $P$ cannot be too high, indicatively
$P=15000$ is a likely upper bound on a modern PC with a 8 GB RAM.

\section{Gauss-Hermite quadrature\label{sec:Gauss-Hermite-quadrature}}

Gauss-Hermite (GHQ) quadrature is a common tool in finance for computing
the expectation of a Gaussian random variable. Here we simply recall
the procedure and we refer the interested reader interested to \citet{abramowitz1964}
and to \citet{judd1998}. 

Let $G\sim\mathcal{N}\left(\mu_{G},\sigma_{G}^{2}\right)$ and $f:\mathbb{R}\rightarrow\mathbb{R}$
a continuous function. In order to compute the expectation $\mathbb{E}\left[f\left(G\right)\right]$,
the GHQ method considers $Q$ possible values $\left\{ g_{q}\right\} _{q=1\dots Q}$
for $G$, with relative weights and the adds together the product
of weight with the images of the points through the function $f$.
Specifically, 
\[
g_{q}=\mu_{G}+\sqrt{2}\sigma_{G}u_{q},
\]
where $\left\{ u_{q}\right\} _{q=1,\dots,Q}$ are the roots of the
Hermite polynomials $H_{Q}$ which occur symmetrically about 0. The
weights $\left\{ w_{q}\right\} _{q=1,\dots,Q}$ are given by the following
expression
\[
w_{i}=\frac{2^{n-1}Q!\sqrt{\pi}}{Q^{2}\left[H_{Q-1}\left(q_{i}\right)\right]^{2}},
\]
then the $Q$-points GHQ quadrature reads

\begin{equation}
\mathbb{E}\left[f\left(G\right)\right]=\frac{1}{\sqrt{\pi}}\int_{-\infty}^{\infty}f(\sqrt{2}\sigma_{G}x+\mu_{G})e^{-x^{2}}dx\approx\frac{1}{\sqrt{\pi}}\sum_{q=1}^{Q}w_{q}f\left(g_{q}\right).
\end{equation}

\section{Proof of Proposition (\ref{prop:1})\label{sec:Proof-of-Proposition}}

First of all, let us point out the linear relation between the two
processes $\mathbf{S}_{n}$ and $\mathbf{B}_{n}$. Specifically, $\mathbf{B}_{n}=\Gamma\mathbf{S}_{n},$
with $\Gamma$ the $d_{n}^{B}\times M$ matrix as follows:{\small{}
\[
\Gamma=\left(\begin{array}{ccccccc}
0 & \frac{1}{M-1} & \dots & \dots & \dots & \dots & \frac{1}{M-1}\\
0 & 0 & \frac{1}{M-2} & \dots & \dots & \dots & \frac{1}{M-2}\\
0 & 0 & 0 & \ddots &  &  & \vdots\\
0 & 0 & 0 & \frac{1}{\min\left\{ n-1,N-M\right\} } & \dots & \dots & \frac{1}{\min\left\{ n-1,N-M\right\} }\\
0 & 0 & 0 & \dots & \dots & 0 & 1
\end{array}\right).
\]
}In order to prove the proposition, it is easier to consider the continuation
value as a function of $\mathbf{S}_{n}$ instead of $\mathbf{B}_{n}$,
so we define 
\[
\mathcal{C}_{n}^{S}\left(\mathbf{S}_{n}\right)=\mathcal{C}_{n}\left(\Gamma\mathbf{S}_{n}\right).
\]
We want to show that $\mathcal{C}_{n}^{S}$ is positively homogeneous
for all $n$ in $\left\{ M,\dots,N\right\} $, that is for all $\kappa>0$
\begin{equation}
\mathcal{C}_{n}^{S}\left(\mathbf{S}_{n}\right)=\kappa\mathcal{C}_{n}^{S}\left(\frac{1}{\kappa}\mathbf{S}_{n}\right).\label{eq:cs}
\end{equation}
The proof is by backward induction. First, we observe that (\ref{eq:cs})
holds for $n=N$ since $\mathcal{C}_{N}^{S}\left(\mathbf{S}_{N}\right)=0$.
Now, we suppose that (\ref{eq:cs}) holds for $n+1$ and we prove
it for $n$. Let $\mathbf{\hat{S}}_{n}=\left(\hat{S}_{t_{n-M+1}},\dots,\hat{S}_{t_{n}}\right)$
be a value for $\mathbf{S}_{n}$. Since we are considering the Black-Scholes
model, we can write 
\[
\left(S_{t_{n+1}}|S_{t_{n}}=\hat{S}_{t_{n}}\right)\sim\hat{S}_{t_{n}}\cdot L,
\]
 with $L=e^{\left(\mu-\frac{\sigma^{2}}{2}\right)\Delta t+\sigma\sqrt{\Delta t}G}$
and $G\sim\mathcal{N}\left(0,1\right)$, that it $L$ has a log-normal
distribution. Then,
\begin{align*}
\mathcal{C}_{n}^{S}\left(\frac{1}{\kappa}\mathbf{\hat{S}}_{n}\right) & =r^{-r\Delta t}\mathbb{E}\left[\max\left(S_{n+1}-\frac{1}{M}\sum_{i=n-M+2}^{n+1}S_{i},\mathcal{C}_{n+1}^{S}\left(\mathbf{S}_{n+1}\right)\right)\left|\mathbf{S}_{n}=\frac{1}{\kappa}\mathbf{\hat{S}}_{n}\right.\right]\\
 & =r^{-r\Delta t}\mathbb{E}\left[\max\left(\left(\frac{M-1}{M}\right)S_{n+1}-\frac{1}{M}\sum_{i=n-M+2}^{n}\frac{\hat{S}_{i}}{\kappa},\mathcal{C}_{n+1}^{S}\left(\frac{\hat{S}_{n-M+2}}{\kappa},\dots,\frac{\hat{S}_{n}}{\kappa},S_{n+1}\right)\right)\left|S_{n}=\frac{1}{\kappa}\hat{S}_{n}\right.\right]\\
 & =re^{-r\Delta t}\mathbb{E}\left[\max\left(\left(\frac{M-1}{M}\right)\frac{1}{\kappa}\hat{S}_{n}L-\frac{1}{M}\sum_{i=n-M+2}^{n}\frac{1}{\kappa}\hat{S}_{i},\mathcal{C}_{n+1}^{S}\left(\frac{1}{\kappa}\hat{S}_{n-M+2},\dots,\frac{1}{\kappa}\hat{S}_{n},\frac{1}{\kappa}\hat{S}_{n}L\right)\right)\right]\\
 & =\frac{1}{k}r^{-r\Delta t}\mathbb{E}\left[\max\left(\left(\frac{M-1}{M}\right)\hat{S}_{n}L-\frac{1}{M}\sum_{i=n-M+2}^{n}\hat{S}_{i},\mathcal{C}_{n+1}^{S}\left(\hat{S}_{n-M+2},\dots,\hat{S}_{n},\hat{S}_{n}L\right)\right)\right]\\
 & =\frac{1}{k}r^{-r\Delta t}\mathbb{E}\left[\max\left(S_{n+1}-\frac{1}{M}\sum_{i=n-M+2}^{n+1}S_{i},\mathcal{C}_{n+1}^{S}\left(\mathbf{S}_{n+1}\right)\right)\left|\mathbf{S}_{n}=\mathbf{\hat{S}}_{n}\right.\right]\\
 & =\frac{1}{k}\mathcal{C}_{n+1}^{S}\left(\mathbf{\hat{S}}_{n}\right),
\end{align*}
which proves (\ref{eq:cs}) since $\mathbf{\hat{S}}_{n}$ is any value
for $\mathbf{S}_{n}$. Then 
\[
\mathcal{C}_{n}\left(\mathbf{B}_{n}\right)=\mathcal{C}_{n}^{S}\left(\mathbf{S}_{n}\right)=\kappa\mathcal{C}_{n}^{S}\left(\frac{1}{\kappa}\mathbf{S}_{n}\right)=\kappa\mathcal{C}_{n}\left(\Gamma\frac{1}{\kappa}\mathbf{S}_{n}\right)=\kappa\mathcal{C}_{n}\left(\frac{1}{\kappa}\Gamma\mathbf{S}_{n}\right)=\kappa\mathcal{C}_{n}\left(\frac{1}{\kappa}\mathbf{B}_{n}\right),
\]
which concludes the proof.
\end{document}